\documentclass[superscriptaddress,twocolumn,preprintnumbers,amsmath,amssymb]{revtex4-2}

\usepackage{graphicx}
\usepackage{dcolumn}
\usepackage{bm}
\usepackage{natbib}
\usepackage{soul}
\usepackage{upgreek}

\makeatletter
\def\@fnsymbol#1{%
   \ifcase#1\or
   \TextOrMath \textdagger \dagger ,  \textasteriskcentered \or 
   \TextOrMath \textdagger \dagger\or
   \TextOrMath\textasteriskcentered *\or
   \TextOrMath \textdaggerdbl \ddagger \or
   \TextOrMath \textsection  \mathsection\or
   \TextOrMath \textparagraph \mathparagraph\or
   \TextOrMath \textbardbl \|\or
   \TextOrMath {\textdagger\textdagger}{\dagger\dagger}\or
   \TextOrMath {\textdaggerdbl\textdaggerdbl}{\ddagger\ddagger}\else
   \@ctrerr \fi
}
\makeatother

\newcommand{\equalcontrib}{These authors contributed equally to this work.}
\newcommand{\corresp}{Corresponding authors. Email:\\ park.jy@skku.edu (J.Y.P.); t.werkmeister@columbia.edu (T.W.); pkim@physics.harvard.edu (P.K.)}
\newcommand{\eqcorresp}{: }

\begin{document}

\title{Vortex-parity-controlled diode effect in Corbino topological Josephson junctions}



\author{Joon Young Park}
\thanks{\eqcorresp}

\affiliation{
Department of Physics, Harvard University, Cambridge, MA 02138, USA
}
\affiliation{
Department of Physics, Sungkyunkwan University (SKKU), Suwon 16419, Republic of Korea
}
\affiliation{
Center for 2D Quantum Heterostructures, Institute for Basic Science (IBS), Sungkyunkwan University (SKKU), Suwon 16419, Republic of Korea
}

\author{Thomas Werkmeister}
\thanks{\eqcorresp}
\affiliation{
John A. Paulson School of Engineering and Applied Sciences, Harvard University, Cambridge, MA 02138, USA
}
\affiliation{
Department of Applied Physics and Applied Mathematics, Columbia University, New York, NY 10027, USA
}
\author{Jonathan Zauberman}
\thanks{\equalcontrib}
\affiliation{
Department of Physics, Harvard University, Cambridge, MA 02138, USA
}
\author{Omri Lesser}
\affiliation{
Department of Physics, Cornell University, Ithaca, New York 14853, USA
}
\affiliation{
Department of Condensed Matter Physics, Weizmann Institute of Science, 7610001 Rehovot, Israel
}
\author{Laurel E. Anderson}
\affiliation{
Department of Physics, Harvard University, Cambridge, MA 02138, USA
}
\author{Yuval Ronen}
\affiliation{
Department of Physics, Harvard University, Cambridge, MA 02138, USA
}
\affiliation{
Department of Condensed Matter Physics, Weizmann Institute of Science, 7610001 Rehovot, Israel
}

\author{Cristian J. Medina Cea}
\affiliation{
Department of Physics, Harvard University, Cambridge, MA 02138, USA
}

\author{Satya K. Kushwaha}
\affiliation{
Department of Chemistry, Princeton University, Princeton, New Jersey 08544, USA
}

\author{Kenji Watanabe}
\affiliation{
Research Center for Electronic and Optical Materials, National Institute for Materials Science, 1-1 Namiki, Tsukuba 305-0044, Japan
}

\author{Takashi Taniguchi}
\affiliation{
Research Center for Materials Nanoarchitectonics, National Institute for Materials Science,  1-1 Namiki, Tsukuba 305-0044, Japan
}

\author{Robert J. Cava}
\affiliation{
Department of Chemistry, Princeton University, Princeton, New Jersey 08544, USA
}
\author{Yuval Oreg}
\affiliation{
Department of Condensed Matter Physics, Weizmann Institute of Science, 7610001 Rehovot, Israel
}
\author{Amir Yacoby}
\affiliation{
Department of Physics, Harvard University, Cambridge, MA 02138, USA
}
\affiliation{
John A. Paulson School of Engineering and Applied Sciences, Harvard University, Cambridge, MA 02138, USA
}
\author{Philip Kim}
\thanks{\corresp}
\affiliation{
Department of Physics, Harvard University, Cambridge, MA 02138, USA
}
\affiliation{
John A. Paulson School of Engineering and Applied Sciences, Harvard University, Cambridge, MA 02138, USA
}

\date{\today}
\begin{abstract}
Nonreciprocal supercurrents in Josephson junctions have recently emerged as a sensitive tool for investigating broken symmetries in superconducting quantum materials. Here, we report an even-odd Josephson diode effect (JDE) in Corbino-geometry junctions fabricated on the pristine surface of a bulk-insulating three-dimensional topological insulator (3DTI). We find that the diode polarity, which indicates the preferred direction of supercurrent flow, robustly alternates its sign depending on the parity (even or odd) of the enclosed vortex number. This behavior is absent in two key control devices: a non-topological graphene Corbino Josephson junction and a 3DTI-based linear Josephson junction. These results indicate that the polarity-tunable JDE is intrinsically linked to the unique combination of the proximitized topological superconductivity in the 3DTI surface and the Corbino device's closed-loop geometry. Our theoretical modeling attributes the observed sign change in diode polarity to the alternating sign of periodic boundary conditions in topological superconductors, supporting the interpretation that the vortex-parity-controlled JDE is a direct manifestation of the underlying Andreev bound state topology associated with the presence of non-Abelian anyons in the vortices.

\end{abstract}

\maketitle

\subsection*{Introduction}

Topological superconductivity has been investigated as a possible platform for fault-tolerant quantum computation, utilizing non-Abelian statistics to provide protected quantum operations \cite{read2000,kitaev2001,ivanov2001,kitaev2003,stern2004,stone2006,nayak2008}. One proposed realization of a topological superconductor is to proximitize the surface states of a three-dimensional topological insulator (3DTI) with an $s$-wave superconductor \cite{fu2008}. Vortices in this system are predicted to host Majorana bound states (MBS) with non-Abelian statistics. One method to localize such vortices with high controllability is to thread magnetic flux through a Josephson junction (JJ) with a 3DTI weak link \cite{grosfeld2011,potter2013}. In geometries where these MBS disperse from zero energy, a $4\pi$-periodic contribution can arise in the current-phase relation (CPR) of the junction in addition to the conventional $2\pi$-periodicity from Cooper pairs \cite{beenakker2013,kurter2015,hegde2020}. Observations of several effects related to this modified CPR have been reported, including skewness of the CPR extracted by asymmetric direct current (DC) superconducting quantum interference device (SQUID) interferometry \cite{kayyalha2020} or scanning SQUID \cite{sochnikov2013,sochnikov2015a}, node lifting in the Fraunhofer pattern of the junctions \cite{kurter2015,yue2024}, and missing Shapiro steps \cite{wiedenmann2016,lecalvez2019,rosenbach2021,rosen2024b}. However, these effects are either inconsistently observed or have possible explanations other than topological superconductivity \cite{lecalvez2019,takeshige2020,dartiailh2021}, including edge effects that can lead to node lifting even for topologically trivial vortices \cite{laubscher2025}. These complications highlight the need for alternative device geometries and robust observables that serve as direct probes of the Andreev bound state (ABS) topology underlying the supercurrent \cite{sauls2018c,prada2020}.

\begin{figure*}[b]
    \centering
    \includegraphics[width = \textwidth]{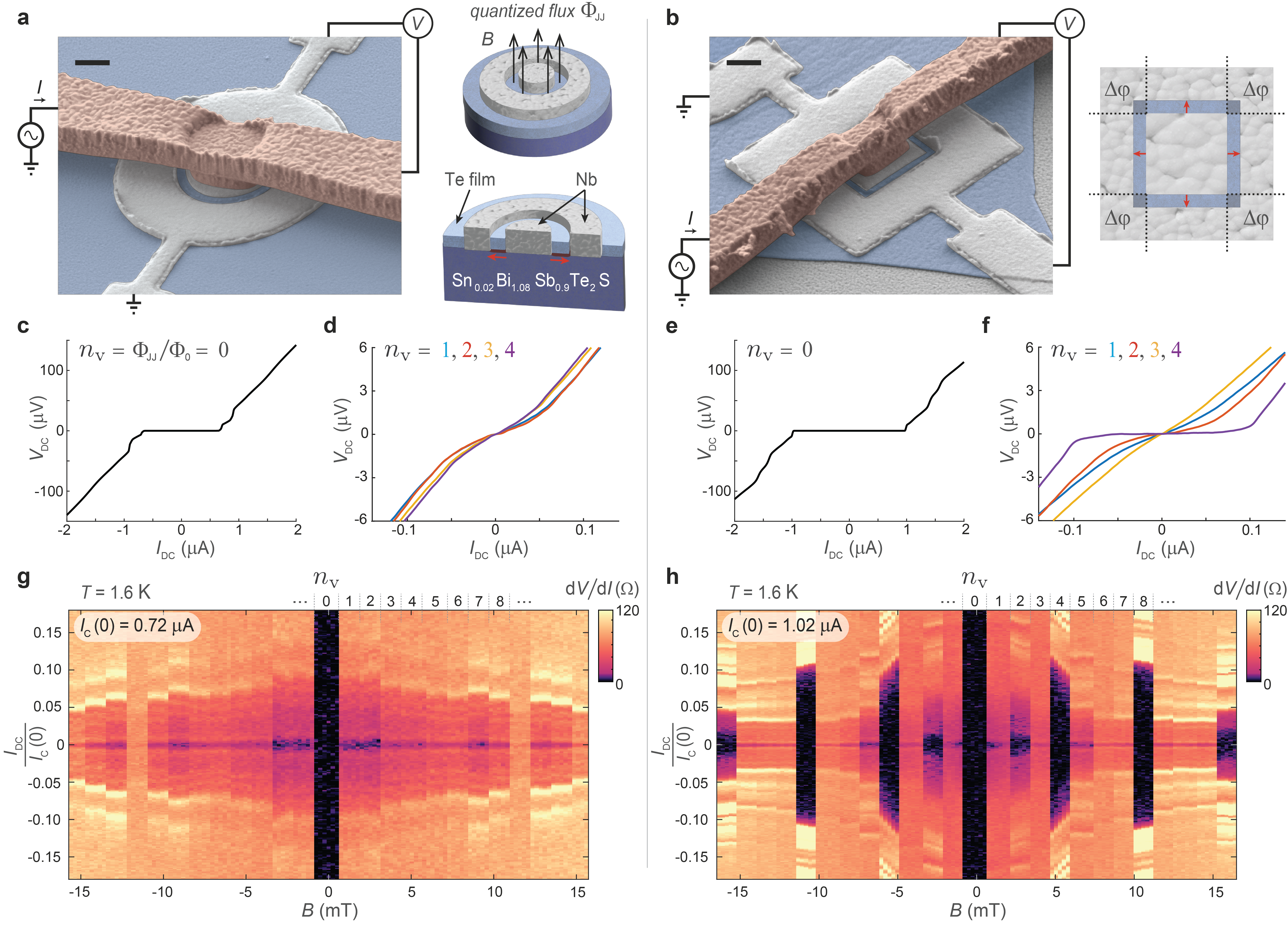}
    \caption{\textbf{Josephson interferometry on circular and square Corbino junctions.}
    \textbf{a}, False-colour scanning electron microscopy (SEM) image of a circular Corbino Josephson junction (scale bar: $300~\text{nm}$). Grey: Au-capped Nb Corbino contacts; light red: Au air bridge; light blue: 3DTI surface with Te capping layer. See Extended Data (ED) Fig. 1 for a large-scale view. The schematic illustrates the quasi-4-probe voltage measurement setup. Upper right: The outer SC contact quantizes magnetic flux $\Phi_{\rm{JJ}}$ into integer multiples of $\Phi_0$. Lower right: Nb directly contacts the 3DTI surface (dark blue) through the milled Te film, with surface states (red) acting as the weak link.
    \textbf{b}, False-colour SEM of a square Corbino junction. The 3DTI flake edges are visible on the Te-coated $\rm{SiO_2/Si}$ substrate (dark grey). Right: Schematic model of the square geometry where four linear junctions are separated by a phase bias $\Delta\varphi$ associated with the corner area, leading to an interference pattern analogous to 4-slit optical diffraction.
    \textbf{c},\textbf{d}, $I_{\rm{DC}}$--$V_{\rm{DC}}$ characteristics of the circular junction at zero flux (\textbf{c}) and at different integer flux quanta (\textbf{d}).
    \textbf{e},\textbf{f}, Same as \textbf{c},\textbf{d} for the square junction.
    \textbf{g}, Differential resistance of the circular junction as a function of $B$ and normalized DC bias current $I_{\rm{DC}}/I_{\rm{c}}(B=0)$. Periodic resistance jumps are visible whenever a vortex enters the weak link region, allowing us to assign an integer $n_{\rm{v}}$ in the junction. A SC (zero-resistance) state is visible only at $n_{\rm{v}}=0$, consistent with a single-slit Fraunhofer diffraction model with discrete sampling at the nodes.
    \textbf{h}, Same as \textbf{g} for the square junction. In agreement with the 4-slit interference model, re-entrant SC features with finite critical currents appear at $n_{\rm v} = \pm 4$, $\pm 8$, $\pm 12$. The additional features at $n_{\rm v} = \pm 2$ arise from the second harmonic of the CPR. Associated with Cooper pair cotunnelling, its doubled phase-winding frequency enables constructive interference at these half-period points where the fundamental mode vanishes.
    All data are taken at $T=1.6~\text{K}$.
    }
    \label{fig1}
\end{figure*}

The edge-free, closed-loop Corbino geometry introduces a unique degree of freedom by imposing periodic boundary conditions that can be altered only in topological systems \cite{fu2008,grosfeld2011,alicea2012,lesser2025}. This geometry also allows for well-separated Josephson vortices within the junction, preventing their loss at open boundaries where the superconducting (SC) gap vanishes \cite{clem2010,grosfeld2011}. Furthermore, it provides a straightforward platform for MBS exchange and braiding, since the angular position of these vortices can be readily controlled by the gauge-invariant phase difference across the junction, $\varphi$ \cite{park2015a,park2020,okugawa2022}. Despite these compelling theoretical advantages, Corbino-geometry JJs (CJJs) have remained relatively unexplored experimentally. Signatures of hysteretic entry and manipulation of vortices~\cite{hadfield2002,hadfield2003} and gate tunability~\cite{matsuo2020} in topologically trivial CJJs have been demonstrated. More recently, control over an integer number of vortices in 3DTI CJJs was demonstrated, although without demonstration of unusual ABS topology~\cite{zhang2022}.

Here, we report the observation of a vortex-parity-controlled Josephson diode effect (JDE) in 3DTI CJJs. Recent studies on Al-InAs and 3DTI linear JJs have explored the JDE, an asymmetry in the critical current with respect to its bias direction, and reported a polarity switching that sometimes coincides with an expected topological phase transition, although this link is not yet conclusive \cite{costa2023,banerjee2023,lotfizadeh2024}. Our central finding is that the JDE polarity in our 3DTI CJJs robustly alternates with the parity (even or odd) of flux quanta trapped in the junction ring. This even-odd effect is absent in a series of control experiments with topologically trivial graphene-based CJJs and 3DTI JJs in a conventional linear geometry, indicating that this effect is unique to the combination of the proximitized 3DTI surface and the closed-loop geometry. Our findings are consistent with numerical models that incorporate the alternating sign of periodic boundary conditions predicted for such a system, pointing to a correspondence between the observed reliance of the JDE on vortex-number parity and the periodic evolution of the topological ABS spectrum. Since each vortex on the 3DTI surface is expected to contain a non-dispersive MBS that determines this periodic evolution, our observation marks an experimental signature of topological superconductivity in our system.

\subsection*{CJJs on single surface of bulk-insulating 3DTI}
To realize high-quality CJJs while preserving the pristine topological surface states of an air-sensitive 3DTI, we have developed a multi-step fabrication process based on an in vacuo protective capping layer and air-bridge contacts (Fig. 1a). We employ single-crystal Sn-doped $\rm{Bi_{1.1}Sb_{0.9}Te_{2}S}$ (Sn-BSTS) as our 3DTI material due to its exceptional properties. The bulk crystal of Sn-BSTS possesses a bulk band gap of $350~\text{meV}$, among the largest of the $\rm{Bi_{2}Te_{3}}$ family 3DTIs, along with a surface state Dirac point energy well-isolated from the bulk bands \cite{kushwaha2016a}. In addition, Sn-BSTS exhibits surface-dominated transport at low temperatures, high electron mobility, and quantum oscillations from the surface states \cite{cai2018}. As a van der Waals (vdW) layered material, the bulk crystal can also be exfoliated to create atomically flat surfaces. While these properties make Sn-BSTS a promising platform for high-quality 3DTI devices, we have found that Sn-BSTS exhibits a greater susceptibility to degradation upon exposure to air than other 3DTIs in the same family (see Supplementary Information (SI) Section I) \cite{kong2011,salehi2015}.

To overcome this challenge, we employ an integrated ultrahigh vacuum (UHV) cluster system where Sn-BSTS single crystals are exfoliated onto $\rm{SiO_2/Si}$ substrates and transferred in vacuo to a molecular beam epitaxy (MBE) module. This allows for the immediate growth of a $10~\text{nm}$ Te thin film without atmospheric exposure, providing a robust protection layer for subsequent fabrication steps. The Te film is insulating at low temperatures, does not perturb the electronic properties of the underlying 3DTI, and grows via vdW epitaxy on $\rm{Bi_{2}Te_{3}}$-family surfaces, ensuring a pristine interface for ex situ processing~\cite{hoefer2015,park2015,okuyama2017,richardson2017,liang2021}. Following electron beam lithography (EBL) of the JJ patterns, the devices are transferred into a process chamber. There, we perform an in situ Ar ion milling to remove the Te film from the contact areas, immediately followed by sputtering of Nb SC contacts. The Nb is then capped with Au in situ to prevent oxidation. The final step is the fabrication of Au air bridges to contact the inner SC electrodes without shorting to other surfaces of the 3DTI. The thickness of the Sn-BSTS flakes used in this study ranges from $20$ to $50~\text{nm}$, which, together with the Corbino geometry, allows for well-isolated single-surface transport on 3DTI Sn-BSTS~\cite{misawa2021}.

\subsection*{Corbino--Josephson interferometry}

We introduce a geometric form of Josephson interferometry that is unique to the Corbino geometry, a technique we term ``Corbino--Josephson interferometry." This method utilizes the junction's shape as a design parameter to characterize the nature of flux entry and selectively resolve the junction's CPR harmonic content, which are crucial prerequisites for our main findings discussed in the subsequent sections. In a CJJ, fluxoid quantization limits the allowed flux in the junction to $\Phi_{\rm JJ} = n_{\rm v}\Phi_0$, assuming no vortices are trapped in the bulk of SC contacts, where $\Phi_0 = h/2e$ is the SC magnetic flux quantum ($h$ is the Planck constant and $e$ is the elementary charge), and $n_{\rm v}$ is the number of vortices. The closed-loop geometry of the CJJ ensures that $n_{\rm v}$ is a well-defined integer determined by the experimental conditions. The general current-phase relation of the JJ can be written
\begin{equation}
    I\left(\varphi\right)=\sum_{k=1}^{\infty}I_{0}^{\left(k\right)}\sin\left(k\varphi\right),
\end{equation}
where $I_{0}^{\left(k\right)}$ is the $k^{\rm th}$ harmonic current and $\varphi$ is the phase difference between the two superconducting electrodes. In a homogeneous circular junction, employing the junction's angular coordinate $\theta$, it can be shown that $\varphi(\theta)=n_{\rm v}\theta+\varphi_0$, where $\varphi_0$ is a global phase offset between the inner and outer SC contacts~\cite{lesser2025}. The critical current $I_{\rm c}$ can be obtained
by maximizing angular averaged $I(\varphi(\theta))$ with respect to $\varphi_0$: 
\begin{equation}
    I_{\rm c}={\rm max}_{\varphi_0}\int_0^{2\pi}\frac{d\theta}{2\pi} \sum_{k=1}^{\infty}I_{0}^{\left(k\right)}\sin\left[k(n_{\rm v}\theta+\varphi_0)\right]
\end{equation}
Due to the circular symmetry, it becomes clear that $I_{\rm c}=0$
for $n_{\rm v}\neq0$, leaving no observable $I_{\rm c}$ at a finite flux.

Breaking the circular symmetry of a CJJ unlocks possibilities for engineering its interference pattern, $I_{\rm c}(\Phi_{\rm JJ})$. For the simplest case of a CJJ with only a first-harmonic CPR, in the shape of a regular polygon with $n_{\rm c}$ corners, this geometric control manifests as a simple selection rule: $I_{\rm c}\neq0$ occurs only when $n_{\rm v}/n_{\rm c}$ is an integer \cite{lesser2025}. This behavior can be understood through a simplified model in which the polygon consists of $n_{\rm c}$ linear junctions on its sides, separated by corners that provide a phase bias, $\Delta\varphi$, associated with corner area (Fig. 1b). This structure effectively acts as a symmetric $n_{\rm c}$-junction, $n_{\rm c}$-loop DC-SQUID, whose interference pattern is analogous to optical $n_{\rm c}$-slit diffraction. The total phase from the enclosed flux, $2\pi n_{\rm v}$, is distributed evenly among these $n_{\rm c}$ units (each comprising a side and a corner), leading to the constructive interference condition $2\pi n_{\rm v}/n_{\rm c} = 2\pi m$, where $m$ is an integer. This geometric Josephson interferometry not only produces a finite $I_{\rm c}$ at nonzero flux but also provides a powerful method to distinguish the harmonic content of the CPR within a single closed-loop device, since the selection rule can be generalized to $I_{\rm c}^{(k)}\neq0$ only when $k n_{\rm v}/n_{\rm c} = m$ (see SI Section II and ref. \cite{lesser2025} for further details).

To demonstrate this concept, we construct 3DTI CJJs in two distinct geometries: circular and square (Fig. 1a,b). The devices are first measured at zero magnetic field to characterize their SC properties. We use a quasi-four-probe configuration where DC bias current $I_{\rm{DC}}$ is sourced to the inner SC electrode via an air bridge and drained from the outer SC electrode, while DC voltage $V_{\rm DC}$ is measured between separate contacts. This setup primarily measures the junction resistance, with a small series contribution ($\ll 1~\Omega$) from the shared portion (the pillar) of the normal metal bridge; we use a normal metal for the bridge, as a SC bridge would trap flux and distort the magnetic field applied for interferometry. Both junctions reach a zero-resistance state below a temperature of $T \leq 3~\text{K}$ (ED Fig. 2). Figure 1c, e show $I_{\rm{DC}}$--$V_{\rm{DC}}$ characteristics at $T=1.6~\text{K}$, where we measure zero-field critical currents $I_{\rm c}(0)$ = $0.72~\upmu\text{A}$ and $1.02~\upmu\text{A}$ for the circular and square CJJs, respectively. They also exhibit finite-voltage features consistent with multiple Andreev reflections, from which we extract an induced SC gap of $\Delta^*\approx0.8~\text{meV}$ (SI Section III). Comparing this gap value with the measured excess current $I_{\rm{e}}$ and normal-state resistance $R_{\rm{N}}$, we apply the Octavio--Tinkham--Blonder--Klapwijk (OTBK) theory (ref. ~\cite{flensberg1988}) to determine a contact transparency of $\mathcal{T} \approx 0.7$, evidencing our fabrication of high-quality 3DTI JJs (ED Table 1).

We then performed interferometry measurements by systematically applying an out-of-plane magnetic field, $B$. To map out the full interference pattern, we employ a ``field-cooling" procedure that ensures one-by-one flux entry. We first warm up the device above the SC transition temperature of the Nb contacts ($T>T_{\rm c, Nb}\approx$ 7~K), set the desired $B$, and then cool the device back to 1.6 K in the presence of the magnetic field. Finally, we sweep $I_{\rm{DC}}$ from zero to positive and then negative values to accurately measure the $I_{\rm c}$ defined as the switching current in both bias directions. This entire cycle is repeated for each value of $B$.

For the circular junction, which serves as a baseline, the behavior is straightforward. A zero differential resistance, $dV_{\rm{DC}}/dI_{\rm{DC}}=0$, with finite $I_{\rm c}$ is observed only in the zero-vortex state ($n_{\rm v}=0$) corresponding to $|B|<\Phi_0/2A$, where $A$ is the effective area enclosed by the outer SC ring (Fig. 1c,d,g). At higher fields, the sequential entry of a single vortex occurs each time the applied flux ($\Phi=BA$) crosses a half-integer multiple of the flux quantum, $(n_{\rm v}+1/2)\Phi_0$. Each periodic entry event is marked by a sharp discontinuity in the otherwise smoothly changing junction resistance, which we use to calculate the effective area of the junction and confirm its correspondence to the geometric area (SI Section IV).

In stark contrast,  the square junction exhibits large re-entrant SC features that can be observed at finite flux, specifically for $n_{\rm v} = \pm 4,\pm 8,\text{and}\pm 12$, with a $I_{\rm c}(B)/I_c(0)$ of order $0.01-0.1$ (Fig. 1f,h). This is a direct consequence of the geometry of the junction, which creates a constructive interference condition for the fundamental harmonic ($k=1$) of the CPR when $n_{\rm v}$ is an integer multiple of $n_{\rm c} = 4$. Furthermore, another nonzero $I_{\rm c}(B)$ is observed at $n_{\rm v}=\pm 2$, arising from the second harmonic ($k=2$) of the CPR, which is expected in our high-transparency junctions \cite{golubov2004}. We note that the absence of re-entrant SC features at $n_{\rm v}=\pm(4m+2)$ for $m\neq0$ is consistent with numerical calculations, which predict a rapid decay of these features for our device's specific corner-to-side ratio \cite{lesser2025}. The well-defined periodicity of both $k=1$ and $k=2$ features, revealed by the Corbino--Josephson interferometry, provides two crucial confirmations for our main discovery: (i) flux enters the junction in a highly controllable, one-by-one manner as Josephson vortices, with no flux being introduced as Abrikosov vortices in the SC contacts, and (ii) our CPR contains significant higher-harmonic content.

\begin{figure*}
    \centering
    \includegraphics[width = \textwidth]{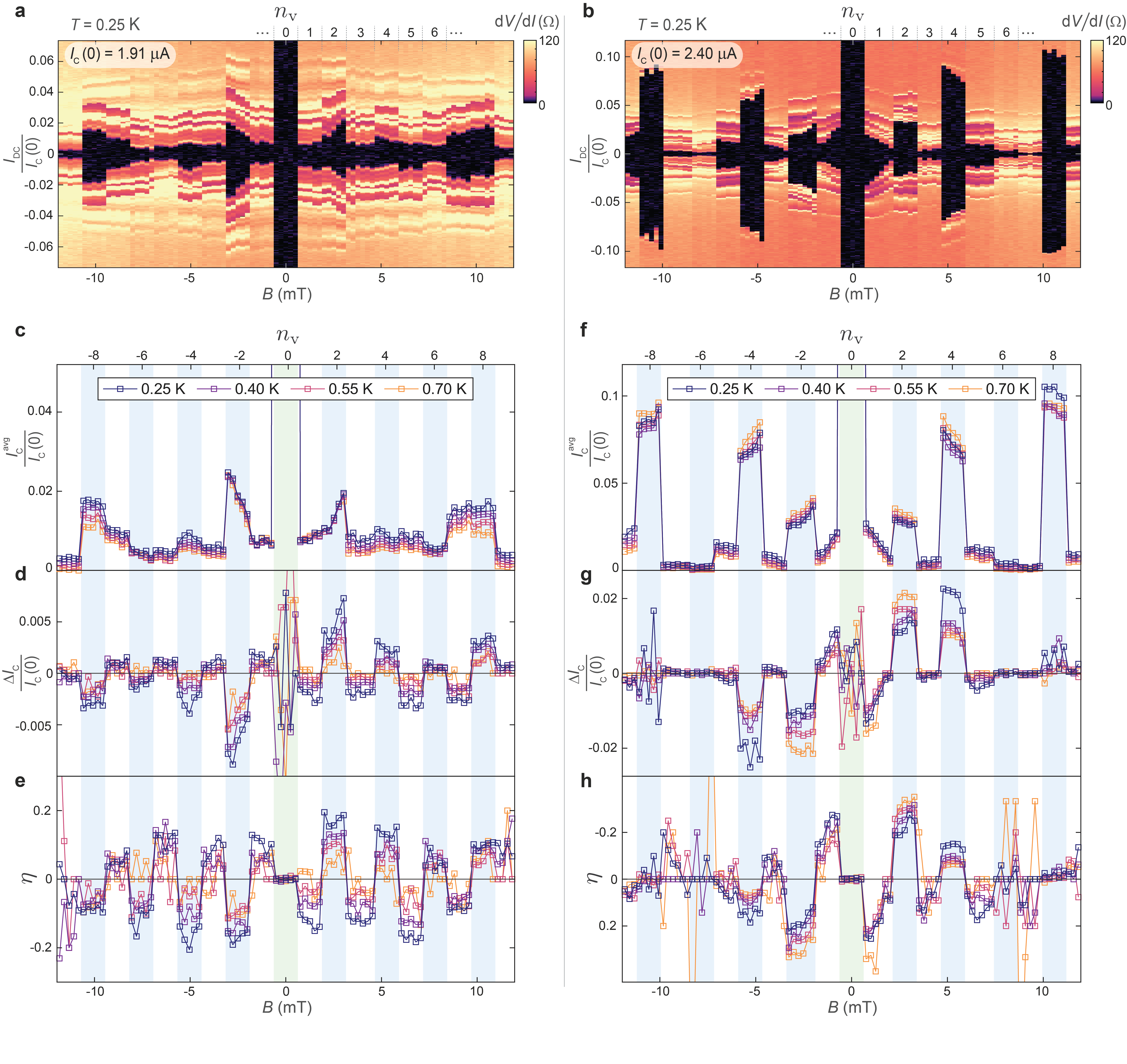}
    \caption{\textbf{Vortex-parity-controlled JDE in 3DTI CJJs.}
    \textbf{a},\textbf{b}, Differential resistance maps of the circular (\textbf{a}) and square (\textbf{b}) junctions as a function of applied magnetic field and bias current normalized by $I_{\rm{c}}(0)$, taken at $T=0.25~\text{K}$. Unlike the 1.6 K data, finite critical currents are resolved across all vortex states.
    \textbf{c}--\textbf{e}, JDE analysis for the circular junction as a function of $B$ at various temperatures:
    \textbf{c}, Average critical current $I_{\rm c}^{\rm avg}(B)=(I_{\rm c}^+(B)+|I_{\rm c}^-(B)|)/2$ normalized by $I_{\rm{c}}(0)$.
    \textbf{d}, Critical current difference $\Delta I_{\rm c}(B)=I_{\rm c}^+(B)-|I_{\rm c}^-(B)|$ normalized by $I_{\rm{c}}(0)$.
    \textbf{e}, Diode efficiency $\eta(B) = \Delta I_{\rm c}(B)/2I_{\rm c}^{\rm avg}(B)$.
    \textbf{f}--\textbf{h}, Same as \textbf{c}--\textbf{e} for the square junction.
    For both devices, the polarity of JDE alternates its sign depending on the parity of the vortex number $n_{\rm{v}}$.
    }
    \label{fig2}
\end{figure*}

\subsection*{Vortex-parity-controlled JDE}

Lowering the field-cooled temperature to $T = 0.25~\text{K}$ causes notable features to emerge within the interference pattern, leading to our key discovery (Fig. 2a,b). The most immediate change is a small but finite residual critical current that is present for all $n_{\rm v}$ (e.g. less than $3\%$ of $I_{\rm c}(0)$ for the circular junction). This residual current is primarily attributed to incomplete destructive interference from junction inhomogeneities, an effect that becomes prominent only as thermal fluctuations are suppressed at lower temperatures. For the square junction, this is likely supplemented by a contribution from an emerging fourth CPR harmonic. This harmonic, negligible at high temperatures due to its small energy scale, always constructively interferes in a square Corbino geometry ($kn_{\rm v}/n_{\rm c} = n_{\rm v}$ for $k=n_{\rm c}=4$), and can thus contribute to the nonzero $I_{\rm c}$ at all values of $n_{\rm v}$.

A closer inspection of this residual critical current reveals a pronounced asymmetry between the positive ($I_{\rm c}^+$) and negative ($I_{\rm c}^-$) critical currents, the definitive signature of the JDE. The emergence of a JDE requires three ingredients: a skewed CPR containing higher harmonic components, broken time-reversal symmetry (TRS), and broken inversion symmetry (IS) \cite{costa2023}. In our experiments, the first two conditions are explicitly met; the interferometry of the previous section confirmed the presence of higher CPR harmonics, and the applied $B$ breaks TRS. The observation of a finite JDE therefore provides strong evidence that IS is also broken. We attribute this primarily to junction inhomogeneities, a point that is corroborated by control experiments below, although the Corbino geometry itself may also play a role. To quantify the JDE, we analyze the normalized average critical current $I_{\rm c}^{\rm avg}(B)/I_{\rm c}(0)=(I_{\rm c}^+(B)+|I_{\rm c}^-(B)|)/2I_{\rm c}(0)$, the normalized critical current difference $\Delta I_{\rm c}(B)/I_{\rm c}(0)=(I_{\rm c}^+(B)-|I_{\rm c}^-(B)|)/I_{\rm c}(0)$ and the diode efficiency $\eta(B) = (I_{\rm c}^+(B)-|I_{\rm c}^-(B)|)/(I_{\rm c}^+(B)+|I_{\rm c}^-(B)|)$ measured at different temperatures.

Fig. 2c--h shows the central observation of this work: a robust, periodic sign reversal of the JDE that is governed by the parity of $n_{\rm v}$. This even-odd effect is most clearly demonstrated in the circular device at $T=0.25~\text{K}\lesssim T_{\rm{c,JJ}}/10$ (see Fig. 2e). Its $\eta$ systematically alternates in sign, being positive (negative) for even (odd) $n_{\rm v}>0$. This alternating pattern vanishes at $n_{\rm v}=0$, consistent with the restoration of effective TRS, and exhibits an overall antisymmetric response satisfying ${\rm sgn}[\eta(-n_{\rm v})]=-{\rm sgn}[\eta(n_{\rm v})]$. This even-odd oscillation is highly robust, persisting up to large vortex numbers ($-8\le n_{\rm v}\le 8$) and at temperatures as high as $T \le0.55~\text{K}$ (see also ED Fig. 3). Remarkably, the effect is not unique to the circular geometry, as we observe the same qualitative behavior in the square device (Fig. 2h). We note that the square junction exhibits a more complex behavior due to its geometry-specific harmonic selection rules; unlike the circular case, different $n_{\rm v}$ states probe different CPR harmonics, and the relative weights of these harmonics vary with temperature. Despite these complications, the even-odd sign reversal in the square junction remains distinct for $-4\le n_{\rm v}\le 5$. Beyond these sign-alternating regimes, $\eta$ for both junctions exhibits irregular $B$- and $T$-dependence. We attribute this to a steeper phase gradient along the junction at larger $n_{\rm v}$ ($\nabla_\theta \varphi \propto n_{\rm v}$), which increases sensitivity to local disorder, consistent with our control devices discussed below. The presence of the vortex-parity-controlled JDE in two distinct geometries thus provides strong evidence that it is an intrinsic and robust feature of our 3DTI CJJs.

The observation of a robust $2\Phi_0$-periodic JDE is particularly striking because its origin cannot be attributed to the ``conventional'' mechanisms for $4\pi$-periodic CPR in topological junctions. Specifically, the constant vortex separation inherent to the CJJ geometry prevents any potential MBS from hybridizing \cite{park2015a,park2020}; therefore, the MBSs expected in our CJJ would be locked at zero energy and not contribute to the supercurrent. Furthermore, our measurement protocol precludes effects that rely on either fermion parity conservation (due to thermal cycling for our field-cooled measurements) or time-sensitive dynamics (as we are in the DC limit). Our results therefore demand an alternative explanation—one that does not rely on MBS dispersing from zero energy, fermion parity conservation, or time-dependent phenomena, yet still produces a robust $2\Phi_0$-periodic observable.

\begin{figure*}
    \centering
    \includegraphics[width = \textwidth]{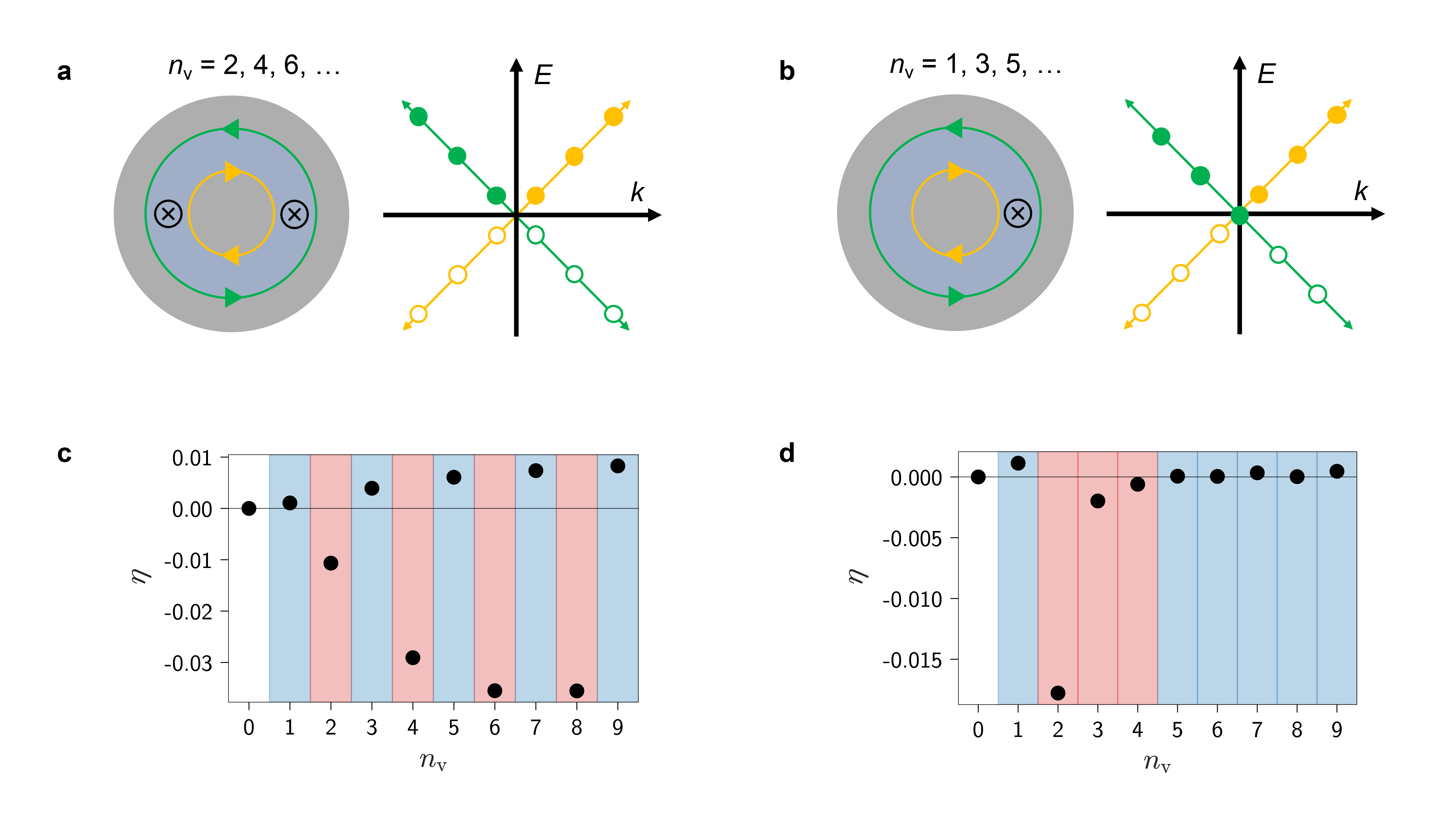}
    \caption{\textbf{Tight-binding model of proximity-induced superconductivity in a 3DTI CJJ.}
    \textbf{a}, Schematic of a 3DTI CJJ. The device is modelled as a proximitized 3DTI weak link (light blue) acting as an effective $p+ip'$ superconductor between Corbino SC contacts (grey). 1D Majorana modes at the edges are denoted in yellow-orange (inner) and green (outer). An even number of magnetic flux quanta threading the junction (black Josephson vortices) leads to anti-periodic boundary conditions for both edge states, restricting their angular momenta $k$ to half-integers. Right: Linear dispersion (colored lines matching the schematic) with allowed states indicated by circles.
    \textbf{b}, Schematic for an odd number of flux quanta. The inner edge state retains half-integer angular momenta, whereas the outer edge state is restricted to integer angular momenta, including a mode at $E,k = 0$.
    \textbf{c}, Calculated Josephson diode efficiency using the topological tight-binding model. The diode polarity (blue: positive, red: negative) alternates with the vortex number parity. Nonzero even vortex numbers correspond to the regime in \textbf{a}, while odd vortex numbers correspond to \textbf{b}.
    \textbf{d}, Calculated diode efficiency using a topologically trivial tight-binding model (schematics not shown). In contrast to the topological case, no clear vortex-parity dependence is observed.
    }
    \label{fig3}
\end{figure*}

\subsection*{Vortex-parity effect in a topological superconductor model}

To shed light on the physical origin of the observed vortex-parity effect, we describe our device using a tight-binding model for a topological superconductor in a Corbino geometry, following Lesser et al.~\cite{lesser2025}. The low-energy physics of our 3DTI CJJs can be described by two counter-propagating one-dimensional Majorana modes, $\gamma_{\rm in}$ at the boundary of the inner superconductor and $\gamma_{\rm out}$ at the inner boundary of the outer superconductor (Fig. 3a,b). The pairing of these two modes then determines the ABS spectrum of the junction from which the Josephson current follows.

The intuitive origin of the observed vortex-parity dependence can be understood by considering the distinct boundary conditions imposed on the individual modes, a feature unique to the closed-loop Corbino geometry. Before considering their pairing, we can analyze these boundary conditions. For an even $n_{\rm v}$, both modes share the same antiperiodic boundary conditions, leading to states with half-integer angular momentum (Fig. 3a) \cite{alicea2012}. In contrast, for an odd $n_{\rm v}$, the outer mode acquires a topological $\pi$ phase shift from encircling the odd flux quanta threaded through the junction. Thus, the presence of an odd number of vortices switches the boundary condition of $\gamma_{\rm out}$ to periodic, shifting its allowed states to integer angular momentum (Fig. 3b). The inner mode's boundary condition, however, always remains antiperiodic. Consequently, the alignment of the two modes' momentum spectra alternates with the parity of $n_{\rm v}$, such that the modes are matched (mismatched) for even (odd) $n_{\rm v}$. This alternating spectral alignment provides the physical foundation for the vortex-parity effect that emerges once pairing is included.

To test this picture numerically, we now include the pairing between the modes, described by the Hamiltonian term $H_\Delta=i\int_0^{2\pi}d\theta \Delta(\theta) \cos\left[\varphi(\theta)/2\right]\gamma_{\rm in}(\theta)\gamma_{\rm out}(\theta)$, where $\varphi(\theta)=n_{\rm v}\theta+\varphi_{0}$ for a circular junction considered here; for non-circular junctions, see ref. \cite{lesser2025} for an extended discussion. To generate a JDE, we also explicitly break IS by introducing a small disordered segment in the ring where the SC pairing potential $\Delta$ deviates from its uniform value. This continuum Hamiltonian can then be discretized and diagonalized numerically~\cite{lesser2025}, yielding eigen-energies from which the CPR and the $\eta$ can be calculated as a function of $n_{\rm v}$.

The calculated $\eta(n_{\rm v})$ for our topological model is shown in Fig. 3c. The numerical results stunningly reproduce our experimental findings, revealing a robust even-odd polarity switching, which we find to be insensitive to the model's microscopic parameters (see SI Section V). This behavior stands in stark contrast to a non-topological control simulation. While a similar model for a non-topological junction can also yield a JDE, the diode polarity computed in this model changes apparently randomly with $n_{\rm v}$ and fails to reproduce the characteristic periodic sign reversal (Fig. 3d). Unlike the topological CJJ, the sign of $\eta(n_{\rm v})$ is instead highly sensitive to the fine-tuning of microscopic parameters, which can be in principle connected to the inhomogeneity in real devices. The strong agreement between our experiment and the topological model, combined with the pronounced disagreement with the non-topological case, supports the interpretation that the vortex-parity-controlled JDE originates from the underlying topological superconductivity.

\subsection*{Absence of vortex-parity-controlled JDE in control devices}

\begin{figure*}
    \centering
    \includegraphics[width = \textwidth]{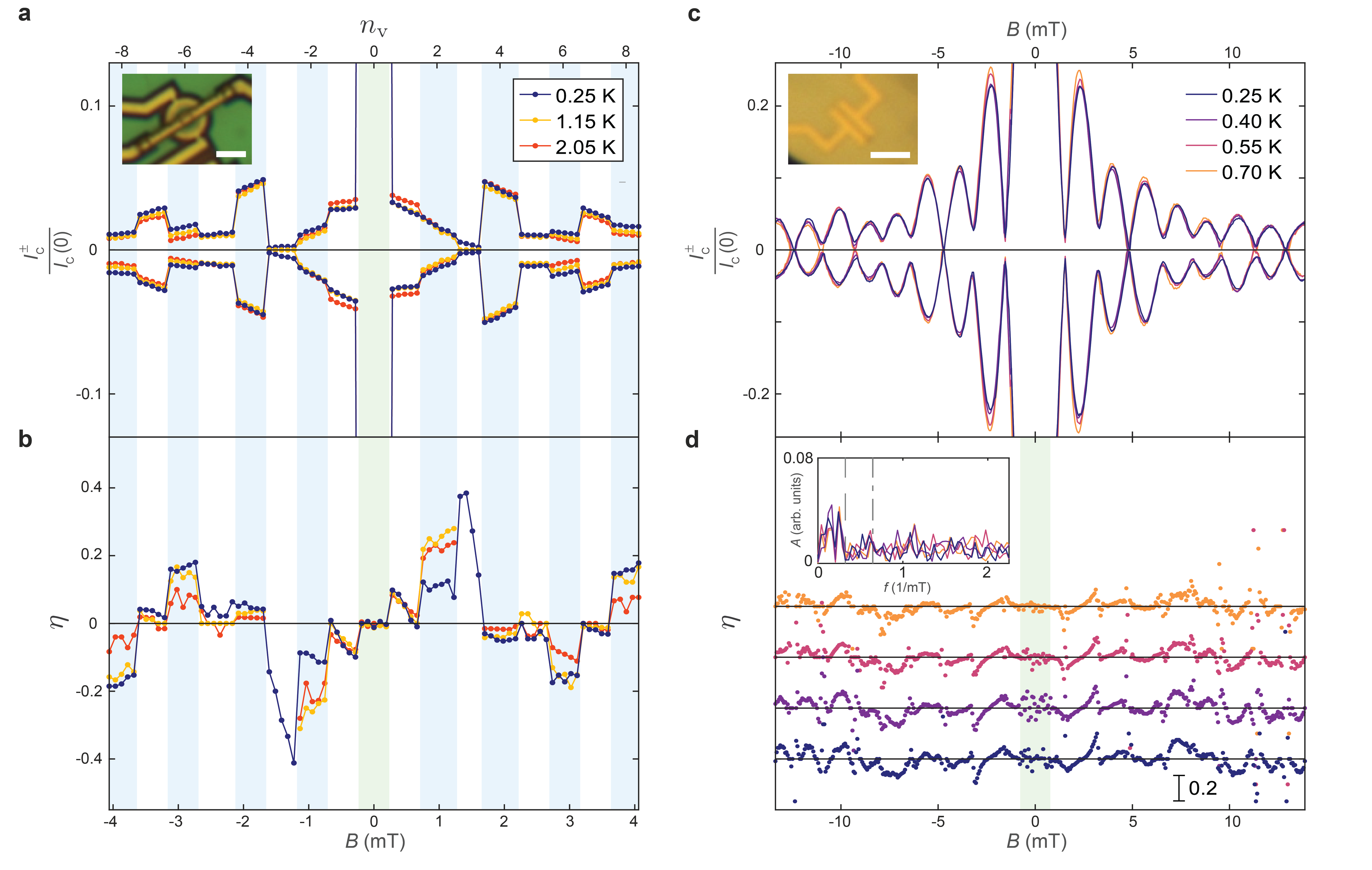}
    \caption{
        \textbf{Absence of vortex-parity-controlled JDE in non-topological graphene Corbino and linear 3DTI JJs.}
        \textbf{a}, Normalized critical current versus applied magnetic field for an electron-doped graphene CJJ, measured at three temperatures. Periodic jumps in $I_{\rm{c}}^{\pm}(B)$ correspond to integer vortex states. Inset: Optical microscopy (OM) image of the device (scale bar: $2~\upmu\text{m}$).
        \textbf{b}, Josephson diode efficiency derived from \textbf{a}. No clear vortex-parity dependence is observed. See ED Fig. 4 for $\eta(B)$ in the extended parameter space.
        \textbf{c}, Normalized critical current versus applied magnetic field for a linear 3DTI JJ, where $\Phi_{\rm{JJ}}$ is tuned continuously, measured at four temperatures. Inset: OM image of the device (scale bar: $2~\upmu\text{m}$).
        \textbf{d}, Josephson diode efficiency derived from \textbf{c}. Curves are offset by $0.4$ for clarity. The inset Fourier transform reveals no clear periodicity. Vertical dash-dotted and dashed lines indicate the major and doubled periodicity of the Fraunhofer pattern in \textbf{c}, respectively; neither matches the frequency components of $\eta(B)$. The central $B$ region (green area, $-0.5 \le \Phi_{\rm{JJ}}/\Phi_{0} < 0.5$) is excluded from the Fourier analysis because the zero-vortex state introduces an additional $\Phi_{0}$ interval between the positive and negative flux branches.
    }
\label{fig4}
\end{figure*}

To experimentally test our model's prediction of an aperiodic JDE in a non-topological system, and to thereby isolate the role of the 3DTI topological surface state, we fabricated a control device using graphene, which has negligible spin-orbit coupling and is thus non-topological \cite{min2006,huertas-hernando2006,yao2007}. Briefly, single-layer graphene is encapsulated by hexagonal boron nitride (hBN), and then side-contacted by molybdenum-rhenium alloy (MoRe) superconductors in a circular Corbino geometry (Fig. 4a inset). This high-transparency junction exhibits a strong JDE with an efficiency reaching 40\% (Fig. 4a,b). Indeed, the diode polarity is highly sensitive to fine-tuning of magnetic flux, temperature, and gate voltage, and shows no discernible even-odd sign alternation anywhere in the wide parameter space we explored (see ED Fig. 4), consistent with the model prediction we discussed above for non-topological CJJs.

To test the importance of the closed-loop geometry, we fabricated a 3DTI JJ in a conventional linear geometry, made concurrently with our main 3DTI CJJ devices (Fig. 4c inset). This open geometry enables continuous flux tuning without thermal cycling, which is expected to produce the well-established sinc-like Fraunhofer interference pattern. Fig.~4c shows the critical current as a function of continuously varied magnetic field in this linear 3DTI JJ. The critical current indeed exhibits a sinc-like Fraunhofer pattern, but with several notable deviations. First, some nodes are lifted from zero critical current. While node lifting can in some cases be tied to a supercurrent originating from MBS hybridization, the stochastic nature of the lifting in our device is more likely attributable to trivial mechanisms such as supercurrent inhomogeneity \cite{yue2024,laubscher2025}. Second, as shown in Fig.~4d, a JDE is also present. Since an ideal linear junction is inversion-symmetric, the mere presence of a JDE provides direct evidence for IS breaking, which we attribute to unavoidable junction inhomogeneities. Despite this, JDE polarity does not show clear periodic switching; its Fourier transform does not reveal a dominant periodicity (Fig.~4d inset).

Taken together, these control experiments underscore the two essential ingredients for the vortex-parity-controlled JDE. The absence of the effect in the graphene device highlights the crucial role of the topological 3DTI material itself. Similarly, its absence in the linear 3DTI junction demonstrates the necessity of the closed-loop Corbino geometry. Our central finding is therefore intrinsically linked to this unique combination of the material's topology and the device's geometry.

\subsection*{Outlook}

In conclusion, we have established the even-odd JDE as a signature whose polarity is locked to the vortex number parity in 3DTI CJJs. We have demonstrated that this effect is intrinsically linked to the unique combination of a topological material and a closed-loop Corbino geometry. Our theoretical model suggests that its origin is a boundary condition, dependent on vortex-parity, imposed upon Majorana edge modes—a mechanism fundamentally distinct from the previous explanations for $4\pi$-periodic phenomena. This work therefore establishes the vortex-parity-controlled JDE not only as a distinct quantum phenomenon but also as a sensitive probe of the underlying topology of Andreev bound state spectrum.

The strong correspondence between our experiment and a model based on topological superconductivity motivates the next crucial steps towards probing non-Abelian anyons, and the Corbino platform presented here offers a tantalizing path towards their manipulation. Encouragingly, control over individual Josephson vortices in Corbino-geometry junctions is expected to provide a direct route to perform braiding operations and unambiguously demonstrate non-Abelian statistics \cite{park2015a,park2020,okugawa2022}, a key requirement for fault-tolerant quantum computation. Our work thus establishes a versatile platform for exploring and controlling the properties of topological superconductors.

\subsection*{Acknowledgments}
We thank C. Strunk, W. D. Oliver, K. Serniak, Y. Ando, G.-H. Lee, S. Park, H.-S. Sim, K. Kim, Y.-J. Doh, K. S. Burch, W. Liu, K. Laubscher, P. Schüffelgen, A. Banerjee, A. Zimmerman, I. Y. Phinney, J. R. Ehrets, and M. E. Wesson for fruitful discussions and help in experiments.
The major part of the experiment was supported by the ONR (N00014-24-1-2081). J.Y.P. acknowledges support from the Institute for Basic Science (IBS-R036-D1) and the National Research Foundation of Korea (NRF) grant funded by the Korean government (MSIT) (No. RS-2021-NR060087). T.W. acknowledges support from the Simons Foundation Society of Fellows (SFI-MPS-SFJ-00011748). Y.R. acknowledges support from the European Research Council Starting Investigator Grant No. 101163917; the Minerva Foundation with funding from the Federal German Ministry for Education and Research; and the Israel Science Foundation (ISF) under Grants Nos. 380/25 and 425/25. The theory part was supported by the European Union’s Horizon 2020 Research and Innovation Programme (Grant Agreement LEGOTOP No. 788715), the DFG (CRC/Transregio 183, EI 519/7-1), the ISF (Grant No. 1914/24), and the ISF–MAFAT program (Grant No. 2478/24). The Sn-BSTS crystals were grown with the support of the Gordon and Betty Moore Foundation grant GBMF-9012. K.W. and T.T. acknowledge support from the JSPS KAKENHI (Grant Nos. 21H05233 and 23H02052), the CREST (JPMJCR24A5), JST and World Premier International Research Center Initiative (WPI), MEXT, Japan. A.Y. was sponsored by the Army Research Office under award W911NF-21-2-0147 and by the Gordon and Betty Moore Foundation grant GBMF 12762. P.K. acknowledges support from AFOSR (8867631-01) for data analysis. Nanofabrication was performed at the Center for Nanoscale Systems at Harvard University, supported in part by an NSF NNIN award ECS-00335765.

\subsection*{Author contributions} 
J.Y.P., T.W., and J.Z. contributed equally. J.Y.P., T.W., Y.R., and P.K. conceived of the experiment. T.W., J.Z., and J.Y.P. performed the nanofabrication, with early contributions from Y.R. and L.E.A. C.J.M.C. contributed to nanofabrication of the graphene control device. J.Y.P., T.W., and J.Z. measured the devices and analyzed the data. O.L. and Y.O. contributed tight-binding calculations and theoretical analysis. S.K.K. and R.J.C. supplied the Sn-BSTS crystals. K.W. and T.T. supplied the hBN crystals. A.Y. and P.K. supervised the project. J.Y.P., T.W., J.Z., and P.K. wrote the paper with inputs from all authors.

\subsection*{Competing interests}
The authors declare no competing interests.

\bibliography{main2}

@article{alicea2012,
  title = {New Directions in the Pursuit of {{Majorana}} Fermions in Solid State Systems},
  author = {Alicea, Jason},
  year = {2012},
  month = jun,
  journal = {Reports on Progress in Physics},
  volume = {75},
  number = {7},
  pages = {076501},
  publisher = {IOP Publishing},
  issn = {0034-4885},
  doi = {10.1088/0034-4885/75/7/076501},
  urldate = {2025-08-20},
  langid = {english}
}

@article{banerjee2023,
  title = {Phase {{Asymmetry}} of {{Andreev Spectra}} from {{Cooper-Pair Momentum}}},
  author = {Banerjee, Abhishek and Geier, Max and Rahman, Md Ahnaf and Thomas, Candice and Wang, Tian and Manfra, Michael J. and Flensberg, Karsten and Marcus, Charles M.},
  year = {2023},
  month = nov,
  journal = {Physical Review Letters},
  volume = {131},
  number = {19},
  pages = {196301},
  publisher = {American Physical Society},
  doi = {10.1103/PhysRevLett.131.196301},
  urldate = {2024-08-27}
}

@article{beenakker2013,
  title = {Search for {{Majorana Fermions}} in {{Superconductors}}},
  author = {Beenakker, C.W.J.},
  year = 2013,
  month = apr,
  journal = {Annual Review of Condensed Matter Physics},
  volume = {4},
  number = {1},
  pages = {113--136},
  issn = {1947-5454, 1947-5462},
  doi = {10.1146/annurev-conmatphys-030212-184337},  
}

@article{cai2018,
  title = {Independence of Topological Surface State and Bulk Conductance in Three-Dimensional Topological Insulators},
  author = {Cai, Shu and Guo, Jing and Sidorov, Vladimir A. and Zhou, Yazhou and Wang, Honghong and Lin, Gongchang and Li, Xiaodong and Li, Yanchuan and Yang, Ke and Li, Aiguo and Wu, Qi and Hu, Jiangping and Kushwaha, Satya K. and Cava, Robert J. and Sun, Liling},
  year = {2018},
  month = nov,
  journal = {npj Quantum Materials},
  volume = {3},
  number = {1},
  pages = {62},
  issn = {2397-4648},
  doi = {10.1038/s41535-018-0134-z}
}

@article{clem2010,
  title = {Corbino-Geometry {{Josephson}} Weak Links in Thin Superconducting Films},
  author = {Clem, John R.},
  year = {2010},
  month = nov,
  journal = {Physical Review B},
  volume = {82},
  number = {17},
  pages = {174515},
  publisher = {American Physical Society},
  doi = {10.1103/PhysRevB.82.174515},
  urldate = {2025-04-04}
}

@article{costa2023,
  title = {Sign Reversal of the {{Josephson}} Inductance Magnetochiral Anisotropy and 0--{$\pi$}-like Transitions in Supercurrent Diodes},
  author = {Costa, A. and Baumgartner, C. and Reinhardt, S. and Berger, J. and Gronin, S. and Gardner, G. C. and Lindemann, T. and Manfra, M. J. and Fabian, J. and Kochan, D. and Paradiso, N. and Strunk, C.},
  year = {2023},
  month = nov,
  journal = {Nature Nanotechnology},
  volume = {18},
  number = {11},
  pages = {1266--1272},
  publisher = {Nature Publishing Group},
  issn = {1748-3395},
  doi = {10.1038/s41565-023-01451-x},
  urldate = {2024-08-27},
  copyright = {2023 The Author(s), under exclusive licence to Springer Nature Limited},
  langid = {english}
}

@article{dartiailh2021,
  title = {Missing {{Shapiro}} Steps in Topologically Trivial {{Josephson}} Junction on {{InAs}} Quantum Well},
  author = {Dartiailh, Matthieu C. and Cuozzo, Joseph J. and Elfeky, Bassel H. and Mayer, William and Yuan, Joseph and Wickramasinghe, Kaushini S. and Rossi, Enrico and Shabani, Javad},
  year = {2021},
  month = jan,
  journal = {Nature Communications},
  volume = {12},
  number = {1},
  pages = {78},
  issn = {2041-1723},
  doi = {10.1038/s41467-020-20382-y}
}

@article{fu2008,
  title = {Superconducting {{Proximity Effect}} and {{Majorana Fermions}} at the {{Surface}} of a {{Topological Insulator}}},
  author = {Fu, Liang and Kane, C. L.},
  year = {2008},
  journal = {Physical Review Letters},
  volume = {100},
  number = {9},
  pages = {096407},
  doi = {10.1103/PhysRevLett.100.096407}
}

@article{golubov2004,
  title = {The Current-Phase Relation in {{Josephson}} Junctions},
  author = {Golubov, A. A. and Kupriyanov, M. {\relax Yu}. and Il'ichev, E.},
  year = {2004},
  month = apr,
  journal = {Reviews of Modern Physics},
  volume = {76},
  number = {2},
  pages = {411--469},
  issn = {0034-6861, 1539-0756},
  doi = {10.1103/RevModPhys.76.411},
  urldate = {2025-08-22},
  copyright = {http://link.aps.org/licenses/aps-default-license},
  langid = {english}
}

@article{grosfeld2011,
  title = {Observing {{Majorana}} Bound States of {{Josephson}} Vortices in Topological Superconductors},
  author = {Grosfeld, Eytan and Stern, Ady},
  year = {2011},
  month = jul,
  journal = {Proceedings of the National Academy of Sciences of the United States of America},
  volume = {108},
  number = {29},
  pages = {11810--4},
  issn = {1091-6490 (Electronic) 0027-8424 (Print) 0027-8424 (Linking)},
  doi = {10.1073/pnas.1101469108},
  pmcid = {PMC3141995}
}

@article{hadfield2002,
  title = {Novel {{Josephson}} Junction Geometries in {{NbCu}} Bilayers Fabricated by Focused Ion Beam Microscope},
  author = {Hadfield, R. H. and Burnell, G. and Grimes, P. K. and Kang, D. J. and Blamire, M. G.},
  year = {2002},
  journal = {Physica C: Superconductivity},
  volume = {367},
  number = {1},
  pages = {267--271},
  issn = {0921-4534},
  doi = {10.1016/S0921-4534(01)01049-8}
}

@article{hadfield2003,
  title = {Corbino Geometry {{Josephson}} Junction},
  author = {Hadfield, Robert H. and Burnell, Gavin and Kang, Dae-Joon and Bell, Chris and Blamire, Mark G.},
  year = {2003},
  journal = {Physical Review B},
  volume = {67},
  number = {14},
  pages = {144513},
  doi = {10.1103/PhysRevB.67.144513}
}

@article{hegde2020,
  title = {A Topological {{Josephson}} Junction Platform for Creating, Manipulating, and Braiding {{Majorana}} Bound States},
  author = {Hegde, Suraj S. and Yue, Guang and Wang, Yuxuan and Huemiller, Erik and Van Harlingen, D. J. and Vishveshwara, Smitha},
  year = {2020},
  journal = {Annals of Physics},
  volume = {423},
  pages = {168326},
  issn = {0003-4916},
  doi = {10.1016/j.aop.2020.168326}
}

@article{hoefer2015,
  title = {Protective Capping of Topological Surface States of Intrinsically Insulating {{Bi}}{\textsubscript{2}}{{Te}}{\textsubscript{3}}},
  author = {Hoefer, Katharina and Becker, Christoph and Wirth, Steffen and Hao Tjeng, Liu},
  year = {2015},
  journal = {AIP Advances},
  volume = {5},
  number = {9},
  pages = {097139},
  issn = {2158-3226},
  doi = {10.1063/1.4931038}
}

@article{huertas-hernando2006,
  title = {Spin-Orbit Coupling in Curved Graphene, Fullerenes, Nanotubes, and Nanotube Caps},
  author = {{Huertas-Hernando}, Daniel and Guinea, F. and Brataas, Arne},
  year = {2006},
  month = oct,
  journal = {Physical Review B},
  volume = {74},
  number = {15},
  pages = {155426},
  publisher = {American Physical Society},
  doi = {10.1103/PhysRevB.74.155426},
  urldate = {2025-08-22}
}

@article{ivanov2001,
  title = {Non-Abelian Statistics of Half-Quantum Vortices in $p$-Wave Superconductors},
  author = {Ivanov, D. A.},
  year = {2001},
  month = jan,
  journal = {Physical Review Letters},
  volume = {86},
  number = {2},
  pages = {268--271},
  publisher = {American Physical Society},
  doi = {10.1103/PhysRevLett.86.268},
  urldate = {2023-10-01}
}

@article{kayyalha2020,
  title = {Highly Skewed Current--Phase Relation in Superconductor--Topological Insulator--Superconductor {{Josephson}} Junctions},
  author = {Kayyalha, Morteza and Kazakov, Aleksandr and Miotkowski, Ireneusz and Khlebnikov, Sergei and Rokhinson, Leonid P. and Chen, Yong P.},
  year = {2020},
  month = jan,
  journal = {npj Quantum Materials},
  volume = {5},
  number = {1},
  pages = {1--7},
  publisher = {Nature Publishing Group},
  issn = {2397-4648},
  doi = {10.1038/s41535-020-0209-5},
  urldate = {2023-09-28},
  copyright = {2020 The Author(s)},
  langid = {english}
}

@article{kitaev2001,
  title = {Unpaired {{Majorana}} Fermions in Quantum Wires},
  author = {Kitaev, A. Yu},
  year = {2001},
  month = oct,
  journal = {Physics-Uspekhi},
  volume = {44},
  number = {10S},
  pages = {131--136},
  publisher = {Uspekhi Fizicheskikh Nauk (UFN) Journal},
  issn = {1063-7869},
  doi = {10.1070/1063-7869/44/10S/S29},
  urldate = {2020-08-29},
  langid = {english}
}

@article{kitaev2003,
  title = {Fault-Tolerant Quantum Computation by Anyons},
  author = {Kitaev, A. {\relax Yu}.},
  year = {2003},
  month = jan,
  journal = {Annals of Physics},
  volume = {303},
  number = {1},
  pages = {2--30},
  issn = {0003-4916},
  doi = {10.1016/S0003-4916(02)00018-0},
  urldate = {2020-08-29},
  langid = {english}
}

@article{kong2011,
  title = {Rapid {{Surface Oxidation}} as a {{Source}} of {{Surface Degradation Factor}} for {{Bi}}{\textsubscript{2}}{{Se}}{\textsubscript{3}}},
  author = {Kong, Desheng and Cha, Judy J. and Lai, Keji and Peng, Hailin and Analytis, James G. and Meister, Stefan and Chen, Yulin and Zhang, Hai-Jun and Fisher, Ian R. and Shen, Zhi-Xun and Cui, Yi},
  year = {2011},
  month = jun,
  journal = {ACS Nano},
  volume = {5},
  number = {6},
  pages = {4698--4703},
  issn = {1936-0851},
  doi = {10.1021/nn200556h}
}

@article{kurter2015,
  title = {Evidence for an Anomalous Current--Phase Relation in Topological Insulator {{Josephson}} Junctions},
  author = {Kurter, C. and Finck, A. D. K. and Hor, Y. S. and Van Harlingen, D. J.},
  year = {2015},
  month = jun,
  journal = {Nature Communications},
  volume = {6},
  number = {1},
  pages = {7130},
  issn = {2041-1723},
  doi = {10.1038/ncomms8130}
}

@article{kushwaha2016a,
  title = {Sn-Doped {{Bi}}{\textsubscript{1.1}}{{Sb}}{\textsubscript{0.9}}{{Te}}{\textsubscript{2}}{{S}} Bulk Crystal Topological Insulator with Excellent Properties},
  author = {Kushwaha, S. K. and Pletikosi{\'c}, I. and Liang, T. and Gyenis, A. and Lapidus, S. H. and Tian, Yao and Zhao, He and Burch, K. S. and Lin, Jingjing and Wang, Wudi and Ji, Huiwen and Fedorov, A. V. and Yazdani, Ali and Ong, N. P. and Valla, T. and Cava, R. J.},
  year = {2016},
  month = apr,
  journal = {Nature Communications},
  volume = {7},
  number = {1},
  pages = {11456},
  issn = {2041-1723},
  doi = {10.1038/ncomms11456}
}

@article{laubscher2025,
  title = {Detection of {{Majorana}} Zero Modes Bound to {{Josephson}} Vortices in Planar Superconductor--Topological Insulator--Superconductor Junctions},
  author = {Laubscher, Katharina and Sau, Jay D.},
  year = 2025,
  month = jun,
  journal = {Physical Review B},
  volume = {111},
  number = {23},
  pages = {235442},
  publisher = {American Physical Society},
  doi = {10.1103/PhysRevB.111.235442},
  urldate = {2025-11-01}
}

@article{lecalvez2019,
  title = {Joule Overheating Poisons the Fractional Ac {{Josephson}} Effect in Topological {{Josephson}} Junctions},
  author = {Le Calvez, K{\'e}vin and Veyrat, Louis and Gay, Fr{\'e}d{\'e}ric and Plaindoux, Philippe and Winkelmann, Clemens B. and Courtois, Herv{\'e} and Sac{\'e}p{\'e}, Benjamin},
  year = {2019},
  month = jan,
  journal = {Communications Physics},
  volume = {2},
  number = {1},
  pages = {1--9},
  publisher = {Nature Publishing Group},
  issn = {2399-3650},
  doi = {10.1038/s42005-018-0100-x},
  urldate = {2023-09-28},
  copyright = {2019 The Author(s)},
  langid = {english}
}

@article{lesser2025,
  title = {Theory of Reentrant Superconductivity in {{Corbino Josephson}} Junctions},
  author = {Lesser, Omri and Park, Joon Young and Ronen, Yuval and Werkmeister, Thomas and Kim, Philip and Oreg, Yuval},
  year = {2026},
  journal = {Preprint}
}

@article{liang2021,
  title = {Protected Long-Time Storage of a Topological Insulator},
  author = {Liang, Luo-Uei and Yen, Yu-Hsiung and Chou, Chia-Wei and Chen, Ko-Hsuan Mandy and Lin, Hsiao-Yu and Huang, Sheng-Wen and Hong, Minghwei and Kwo, Jueinai and Hoffmann, Germar},
  year = {2021},
  journal = {AIP Advances},
  volume = {11},
  number = {2},
  pages = {025245},
  issn = {2158-3226},
  doi = {10.1063/5.0037751}
}

@article{lotfizadeh2024,
  title = {Superconducting Diode Effect Sign Change in Epitaxial {{Al-InAs Josephson}} Junctions},
  author = {Lotfizadeh, Neda and Schiela, William F. and Pekerten, Bar{\i}{\c s} and Yu, Peng and Elfeky, Bassel Heiba and Strickland, William M. and {Matos-Abiague}, Alex and Shabani, Javad},
  year = {2024},
  month = apr,
  journal = {Communications Physics},
  volume = {7},
  number = {1},
  pages = {1--8},
  publisher = {Nature Publishing Group},
  issn = {2399-3650},
  doi = {10.1038/s42005-024-01618-5},
  urldate = {2024-08-27},
  copyright = {2024 The Author(s)},
  langid = {english}
}

@article{matsuo2020,
  title = {Evaluation of the Vortex Core Size in Gate-Tunable {{Josephson}} Junctions in {{Corbino}} Geometry},
  author = {Matsuo, Sadashige and Tateno, Mizuki and Sato, Yosuke and Ueda, Kento and Takeshige, Yuusuke and Kamata, Hiroshi and Lee, Joon Sue and Shojaei, Borzoyeh and Palmstr{\o}m, Christopher J. and Tarucha, Seigo},
  year = {2020},
  journal = {Physical Review B},
  volume = {102},
  number = {4},
  pages = {045301},
  doi = {10.1103/PhysRevB.102.045301}
}

@article{min2006,
  title = {Intrinsic and {{Rashba}} Spin-Orbit Interactions in Graphene Sheets},
  author = {Min, Hongki and Hill, J. E. and Sinitsyn, N. A. and Sahu, B. R. and Kleinman, Leonard and MacDonald, A. H.},
  year = {2006},
  month = oct,
  journal = {Physical Review B},
  volume = {74},
  number = {16},
  pages = {165310},
  issn = {1098-0121, 1550-235X},
  doi = {10.1103/PhysRevB.74.165310},
  urldate = {2025-08-22},
  copyright = {http://link.aps.org/licenses/aps-default-license},
  langid = {english}
}

@article{misawa2021,
  title = {Single-Surface Conduction in a Highly Bulk-Resistive Topological Insulator {{Sn}}{\textsubscript{0.02}}{{Bi}}{\textsubscript{1.08}}{{Sb}}{\textsubscript{0.9}}{{Te}}{\textsubscript{2}}{{S}} Using the {{Corbino}} Geometry},
  author = {Misawa, Tetsuro and Nakamura, Shuji and Okazaki, Yuma and Fukuyama, Yasuhiro and Nasaka, Nariaki and Ezure, Hiroki and Urano, Chiharu and Kaneko, Nobu-Hisa and Sasagawa, Takao},
  year = {2021},
  journal = {Applied Physics Letters},
  volume = {118},
  number = {3},
  pages = {033102},
  issn = {0003-6951},
  doi = {10.1063/5.0026730}
}

@article{nayak2008,
  title = {Non-{{Abelian}} Anyons and Topological Quantum Computation},
  author = {Nayak, Chetan and Simon, Steven H. and Stern, Ady and Freedman, Michael and Das Sarma, Sankar},
  year = {2008},
  month = sep,
  journal = {Reviews of Modern Physics},
  volume = {80},
  number = {3},
  pages = {1083--1159},
  publisher = {American Physical Society},
  doi = {10.1103/RevModPhys.80.1083},
  urldate = {2020-08-29}
}

@article{okugawa2022,
  title = {Vortex Control in Superconducting {{Corbino}} Geometry Networks},
  author = {Okugawa, T. and Park, S. and Recher, P. and Kennes, D. M.},
  year = {2022},
  journal = {Physical Review B},
  volume = {106},
  number = {2},
  pages = {024501},
  doi = {10.1103/PhysRevB.106.024501}
}

@article{okuyama2017,
  title = {Growth and Atomic Structure of Tellurium Thin Films Grown on {{Bi}}{\textsubscript{2}}{{Te}}{\textsubscript{3}}},
  author = {Okuyama, Yuma and Sugiyama, Yuya and Ideta, Shin-ichiro and Tanaka, Kiyohisa and Hirahara, Toru},
  year = {2017},
  journal = {Applied Surface Science},
  volume = {398},
  pages = {125--129},
  issn = {0169-4332},
  doi = {10.1016/j.apsusc.2016.11.196}
}

@article{park2015,
  title = {Crystallinity of Tellurium Capping and Epitaxy of Ferromagnetic Topological Insulator Films on {{SrTiO}}{\textsubscript{3}}},
  author = {Park, Jihwey and Soh, Yeong-Ah and Aeppli, Gabriel and Feng, Xiao and Ou, Yunbo and He, Ke and Xue, Qi-Kun},
  year = {2015},
  month = jun,
  journal = {Scientific Reports},
  volume = {5},
  number = {1},
  pages = {11595},
  issn = {2045-2322},
  doi = {10.1038/srep11595}
}

@article{park2015a,
  title = {Detecting the {{Exchange Phase}} of {{Majorana Bound States}} in a {{Corbino Geometry Topological Josephson Junction}}},
  author = {Park, Sunghun and Recher, Patrik},
  year = {2015},
  journal = {Physical Review Letters},
  volume = {115},
  number = {24},
  pages = {246403},
  doi = {10.1103/PhysRevLett.115.246403}
}

@article{park2020,
  title = {Electron-{{Tunneling-Assisted Non-Abelian Braiding}} of {{Rotating Majorana Bound States}}},
  author = {Park, Sunghun and Sim, H.-S. and Recher, Patrik},
  year = {2020},
  month = oct,
  journal = {Physical Review Letters},
  volume = {125},
  number = {18},
  pages = {187702},
  issn = {0031-9007, 1079-7114},
  doi = {10.1103/PhysRevLett.125.187702},
  urldate = {2023-10-02},
  langid = {english}
}

@article{potter2013,
  title = {Anomalous Supercurrent from {{Majorana}} States in Topological Insulator {{Josephson}} Junctions},
  author = {Potter, Andrew C. and Fu, Liang},
  year = {2013},
  journal = {Physical Review B},
  volume = {88},
  number = {12},
  pages = {121109},
  doi = {10.1103/PhysRevB.88.121109}
}

@article{prada2020,
  title = {From {{Andreev}} to {{Majorana}} Bound States in Hybrid Superconductor--Semiconductor Nanowires},
  author = {Prada, Elsa and {San-Jose}, Pablo and {de Moor}, Michiel W. A. and Geresdi, Attila and Lee, Eduardo J. H. and Klinovaja, Jelena and Loss, Daniel and Nyg{\aa}rd, Jesper and Aguado, Ram{\'o}n and Kouwenhoven, Leo P.},
  year = 2020,
  month = oct,
  journal = {Nature Reviews Physics},
  volume = {2},
  number = {10},
  pages = {575--594},
  publisher = {Nature Publishing Group},
  issn = {2522-5820},
  doi = {10.1038/s42254-020-0228-y}
}

@article{read2000,
  title = {Paired States of Fermions in Two Dimensions with Breaking of Parity and Time-Reversal Symmetries and the Fractional Quantum {{Hall}} Effect},
  author = {Read, N. and Green, Dmitry},
  year = {2000},
  month = apr,
  journal = {Physical Review B},
  volume = {61},
  number = {15},
  pages = {10267--10297},
  publisher = {American Physical Society},
  doi = {10.1103/PhysRevB.61.10267},
  urldate = {2023-10-01}
}

@article{richardson2017,
  title = {Structural Properties of Thin-Film Ferromagnetic Topological Insulators},
  author = {Richardson, C. L. and {Devine-Stoneman}, J. M. and Divitini, G. and Vickers, M. E. and Chang, C. Z. and Amado, M. and Moodera, J. S. and Robinson, J. W. A.},
  year = {2017},
  month = sep,
  journal = {Scientific Reports},
  volume = {7},
  number = {1},
  pages = {12061},
  issn = {2045-2322},
  doi = {10.1038/s41598-017-12237-2}
}

@article{ronen2021,
  title = {Aharonov--{{Bohm}} Effect in Graphene-Based {{Fabry}}--{{P{\'e}rot}} Quantum {{Hall}} Interferometers},
  author = {Ronen, Yuval and Werkmeister, Thomas and Haie Najafabadi, Danial and Pierce, Andrew T. and Anderson, Laurel E. and Shin, Young Jae and Lee, Si Young and Lee, Young Hee and Johnson, Bobae and Watanabe, Kenji and Taniguchi, Takashi and Yacoby, Amir and Kim, Philip},
  year = {2021},
  month = may,
  journal = {Nature Nanotechnology},
  volume = {16},
  number = {5},
  pages = {563--569},
  issn = {1748-3395},
  doi = {10.1038/s41565-021-00861-z}
}

@article{rosen2024b,
  title = {Fractional {{AC Josephson}} Effect in a Topological Insulator Proximitized by a Self-Formed Superconductor},
  author = {Rosen, Ilan T. and Trimble, Christie J. and Andersen, Molly P. and Mikheev, Evgeny and Li, Yanbin and Liu, Yunzhi and Tai, Lixuan and Zhang, Peng and Wang, Kang L. and Cui, Yi and Kastner, M. A. and Williams, James R. and {Goldhaber-Gordon}, David},
  year = {2024},
  month = aug,
  journal = {Physical Review B},
  volume = {110},
  number = {6},
  pages = {064511},
  publisher = {American Physical Society},
  doi = {10.1103/PhysRevB.110.064511},
  urldate = {2025-04-03}
}

@article{rosenbach2021,
  title = {Reappearance of First {{Shapiro}} Step in Narrow Topological {{Josephson}} Junctions},
  author = {Rosenbach, Daniel and Schmitt, Tobias W. and Sch{\"u}ffelgen, Peter and Stehno, Martin P. and Li, Chuan and Schleenvoigt, Michael and Jalil, Abdur R. and Mussler, Gregor and Neumann, Elmar and Trellenkamp, Stefan and Golubov, Alexander A. and Brinkman, Alexander and Gr{\"u}tzmacher, Detlev and Sch{\"a}pers, Thomas},
  year = {2021},
  month = jun,
  journal = {Science Advances},
  volume = {7},
  number = {26},
  pages = {eabf1854},
  publisher = {American Association for the Advancement of Science},
  doi = {10.1126/sciadv.abf1854},
  urldate = {2025-04-03}
}

@article{salehi2015,
  title = {Stability of Low-Carrier-Density Topological-Insulator {{Bi}}{\textsubscript{2}}{{Se}}{\textsubscript{3}} Thin Films and Effect of Capping Layers},
  author = {Salehi, Maryam and Brahlek, Matthew and Koirala, Nikesh and Moon, Jisoo and Wu, Liang and Armitage, N. P. and Oh, Seongshik},
  year = {2015},
  journal = {APL Materials},
  volume = {3},
  number = {9},
  pages = {091101},
  issn = {2166-532X},
  doi = {10.1063/1.4931767}
}

@article{sauls2018c,
  title = {Andreev Bound States and Their Signatures},
  author = {Sauls, J. A.},
  year = 2018,
  month = jun,
  journal = {Philosophical Transactions of the Royal Society A: Mathematical, Physical and Engineering Sciences},
  volume = {376},
  number = {2125},
  pages = {20180140},
  issn = {1364-503X},
  doi = {10.1098/rsta.2018.0140}
}

@article{sochnikov2013,
  title = {Direct {{Measurement}} of {{Current-Phase Relations}} in {{Superconductor}}/{{Topological Insulator}}/{{Superconductor Junctions}}},
  author = {Sochnikov, Ilya and Bestwick, Andrew J. and Williams, James R. and Lippman, Thomas M. and Fisher, Ian R. and {Goldhaber-Gordon}, David and Kirtley, John R. and Moler, Kathryn A.},
  year = {2013},
  month = jul,
  journal = {Nano Letters},
  volume = {13},
  number = {7},
  pages = {3086--3092},
  publisher = {American Chemical Society},
  issn = {1530-6984},
  doi = {10.1021/nl400997k},
  urldate = {2025-04-03}
}

@article{sochnikov2015a,
  title = {Nonsinusoidal {{Current-Phase Relationship}} in {{Josephson Junctions}} from the {{3D Topological Insulator HgTe}}},
  author = {Sochnikov, Ilya and Maier, Luis and Watson, Christopher A. and Kirtley, John R. and Gould, Charles and Tkachov, Grigory and Hankiewicz, Ewelina M. and Br{\"u}ne, Christoph and Buhmann, Hartmut and Molenkamp, Laurens W. and Moler, Kathryn A.},
  year = {2015},
  month = feb,
  journal = {Physical Review Letters},
  volume = {114},
  number = {6},
  pages = {066801},
  publisher = {American Physical Society},
  doi = {10.1103/PhysRevLett.114.066801},
  urldate = {2025-04-03}
}

@article{stern2004,
  title = {Geometric Phases and Quantum Entanglement as Building Blocks for Non-{{Abelian}} Quasiparticle Statistics},
  author = {Stern, Ady and {von Oppen}, Felix and Mariani, Eros},
  year = {2004},
  month = nov,
  journal = {Physical Review B},
  volume = {70},
  number = {20},
  pages = {205338},
  publisher = {American Physical Society},
  doi = {10.1103/PhysRevB.70.205338},
  urldate = {2023-10-01}
}

@article{stone2006,
  title = {Fusion Rules and Vortices in $p_x+ip_y$ Superconductors},
  author = {Stone, Michael and Chung, Suk-Bum},
  year = {2006},
  month = jan,
  journal = {Physical Review B},
  volume = {73},
  number = {1},
  pages = {014505},
  publisher = {American Physical Society},
  doi = {10.1103/PhysRevB.73.014505},
  urldate = {2023-10-01}
}

@article{takeshige2020,
  title = {Experimental Study of Ac Josephson Effect in Gate-Tunable $(\text{Bi}_{1-x}\text{Sb}_x)_2\text{Te}_3$ Thin-Film Josephson Junctions},
  author = {Takeshige, Yuusuke and Matsuo, Sadashige and Deacon, Russell S. and Ueda, Kento and Sato, Yosuke and Zhao, Yi-Fan and Zhou, Lingjie and Chang, Cui-Zu and Ishibashi, Koji and Tarucha, Seigo},
  year = {2020},
  month = mar,
  journal = {Physical Review B},
  volume = {101},
  number = {11},
  pages = {115410},
  publisher = {American Physical Society},
  doi = {10.1103/PhysRevB.101.115410},
  urldate = {2023-09-28}
}

@article{wiedenmann2016,
  title = {4{$\pi$}-Periodic {{Josephson}} Supercurrent in {{HgTe-based}} Topological {{Josephson}} Junctions},
  author = {Wiedenmann, J. and Bocquillon, E. and Deacon, R. S. and Hartinger, S. and Herrmann, O. and Klapwijk, T. M. and Maier, L. and Ames, C. and Br{\"u}ne, C. and Gould, C. and Oiwa, A. and Ishibashi, K. and Tarucha, S. and Buhmann, H. and Molenkamp, L. W.},
  year = {2016},
  month = jan,
  journal = {Nature Communications},
  volume = {7},
  number = {1},
  pages = {10303},
  issn = {2041-1723},
  doi = {10.1038/ncomms10303}
}

@phdthesis{werkmeister2025,
author={Werkmeister,Thomas},
year={2025},
school= {Harvard University},
title={Interferometry of Integer and Fractional Quantum Hall Edge States in Graphene},
journal={ProQuest Dissertations and Theses},
pages={190},
isbn={9798280713758},
url={https://ezproxy.cul.columbia.edu/login?url=https://www.proquest.com/dissertations-theses/interferometry-integer-fractional-quantum-hall/docview/3217394627/se-2},
}

@article{yao2007,
  title = {Spin-Orbit Gap of Graphene: {{First-principles}} Calculations},
  shorttitle = {Spin-Orbit Gap of Graphene},
  author = {Yao, Yugui and Ye, Fei and Qi, Xiao-Liang and Zhang, Shou-Cheng and Fang, Zhong},
  year = {2007},
  month = jan,
  journal = {Physical Review B},
  volume = {75},
  number = {4},
  pages = {041401},
  issn = {1098-0121, 1550-235X},
  doi = {10.1103/PhysRevB.75.041401},
  urldate = {2025-08-22},
  copyright = {http://link.aps.org/licenses/aps-default-license},
  langid = {english}
}

@article{yue2024,
  title = {Signatures of {{Majorana}} Bound States in the Diffraction Patterns of Extended Superconductor--Topological Insulator--Superconductor {{Josephson}} Junctions},
  author = {Yue, Guang and Zhang, Can and Huemiller, Erik D. and Montone, Jessica H. and Arias, Gilbert R. and Wild, Drew G. and Zhang, Jered Y. and Hamilton, David R. and Yuan, Xiaoyu and Yao, Xiong and Jain, Deepti and Moon, Jisoo and Salehi, Maryam and Koirala, Nikesh and Oh, Seongshik and Van Harlingen, Dale J.},
  year = {2024},
  month = mar,
  journal = {Physical Review B},
  volume = {109},
  number = {9},
  pages = {094511},
  publisher = {American Physical Society},
  doi = {10.1103/PhysRevB.109.094511},
  urldate = {2025-04-03}
}

@article{zhang2022,
  title = {Ac {{Josephson}} Effect in {{Corbino-geometry Josephson}} Junctions Constructed on {{Bi}}{\textsubscript{2}}{{Te}}{\textsubscript{3}} Surface},
  author = {Zhang, Yunxiao and Lyu, Zhaozheng and Wang, Xiang and Zhuo, Enna and Sun, Xiaopei and Li, Bing and Shen, Jie and Liu, Guangtong and Qu, Fanming and L{\"u}, Li},
  year = {2022},
  month = oct,
  journal = {Chinese Physics B},
  volume = {31},
  number = {10},
  pages = {107402},
  issn = {1674-1056},
  doi = {10.1088/1674-1056/ac89d4}
}

@article{flensberg1988,
  title = {Subharmonic Energy-Gap Structure in Superconducting Weak Links},
  author = {Flensberg, K.},
  year = {1988},
  journal = {Physical Review B},
  volume = {38},
  number = {13},
  pages = {8707--8711},
  doi = {10.1103/PhysRevB.38.8707}
}

@article{schmitt2022,
  title = {Anomalous Temperature Dependence of Multiple {{Andreev}} Reflections in a Topological Insulator {{Josephson}} Junction},
  author = {Schmitt, Tobias W and Frohn, Benedikt and Wittl, Wilhelm and Jalil, Abdur R and Schleenvoigt, Michael and Zimmermann, Erik and Schmidt, Anne and Sch{\"a}pers, Thomas and Cuevas, Juan Carlos and Brinkman, Alexander and Gr{\"u}tzmacher, Detlev and Sch{\"u}ffelgen, Peter},
  year = {2022},
  month = dec,
  journal = {Superconductor Science and Technology},
  volume = {36},
  number = {2},
  pages = {024002},
  publisher = {IOP Publishing},
  issn = {0953-2048},
  doi = {10.1088/1361-6668/aca4fe},
  urldate = {2025-10-15},
  langid = {english}
}

@article{gubin2005,
  title = {Dependence of Magnetic Penetration Depth on the Thickness of Superconducting {{Nb}} Thin Films},
  author = {Gubin, A. I. and Il'in, K. S. and Vitusevich, S. A. and Siegel, M. and Klein, N.},
  year = 2005,
  month = aug,
  journal = {Physical Review B},
  volume = {72},
  number = {6},
  pages = {064503},
  issn = {1098-0121, 1550-235X},
  doi = {10.1103/PhysRevB.72.064503},
  urldate = {2025-10-30},
  copyright = {http://link.aps.org/licenses/aps-default-license},
  langid = {english}
}


\subsection*{Methods}

\subsubsection*{Device fabrication}
The fabrication of 3DTI JJs was performed using the integrated UHV cluster system described in the main text, which connects an ultrahigh-purity Ar glovebox (MBRAUN; frost point $\approx-100^\circ \text{C}$, equivalent to $\approx 10$ parts per billion (ppb) of $\rm{H_2O}$ by volume), a UHV MBE module (Effucell; base pressure $<10^{-10}~\text{Torr}$), and dedicated chambers for surface preparation and multi-process deposition (AJA International). First, to eliminate moisture and surface contaminants, the $\text{SiO}_2/\text{Si}$ substrate underwent a prolonged bake prior to plasma cleaning in the connected preparation chamber. A bulk Sn-BSTS single crystal, preserved at $-35^\circ\text{C}$ within the glovebox, was mechanically exfoliated onto the clean substrate under the same inert environment and subsequently transferred in vacuo to the MBE chamber. A $10~\text{nm}$ Te thin film was deposited to protect the pristine Sn-BSTS surface during subsequent atmospheric exposure required for lithography.

Candidate flakes were identified using OM and atomic force microscopy based on large, atomically flat terraces sufficient to accommodate the entire Corbino footprint, suitable thicknesses, and lateral geometries optimized for air bridge integration. EBL was performed using a bilayer poly(methyl methacrylate) (PMMA) resist process to define the Corbino inner and outer contacts as well as external leads. Each PMMA layer was baked in high vacuum at $80^\circ \text{C}$ to minimize degradation of the Sn-BSTS flakes. The sample was then loaded into the cluster's deposition chamber, followed by in situ Ar ion milling to remove the $10~\text{nm}$ Te film and etch minimally ($\approx 1\text{--}3~\text{nm}$) into the Sn-BSTS surface. Immediately following the milling, a $50~\text{nm}$-thick Nb film was deposited via DC magnetron sputtering, capped in situ with $20~\text{nm}$ of Au via electron-beam evaporation. Lift-off was performed in $50^\circ \text{C}$ acetone, aided by ultrasonication to ensure a clean definition of the narrow channel (length $L = 110 \pm 10~\text{nm}$, estimated by SEM). The Au capping layer prevents surface oxidation of the Nb; due to the short superconducting coherence length of Nb, the normal metal cap has a minimal adverse effect on the superconducting properties. Air bridges were defined by another EBL step using PMMA/copolymer bilayer resist, followed by thermal evaporation of a thick normal metal layer ($5~\text{nm}$ Cr / $350~\text{nm}$ Au), as described in previous works~\cite{ronen2021,werkmeister2025}.

For the control graphene CJJs, hBN/graphene/hBN heterostructures were prepared using the standard dry-transfer method, where stamps consisting of polydimethylsiloxane (PDMS) coated with a polycarbonate (PC) polymer were used to pick up each flake sequentially. The stack was then etched using reactive ion etching following an EBL step to define and isolate multiple devices. The graphene CJJ channel had a nominal length of $L = 400~\text{nm}$. An additional EBL step was performed to define junction contacts and leads, followed by DC magnetron sputtering of $50~\text{nm}$ MoRe and electron-beam deposition of a $20~\text{nm}$ Au capping layer, forming superconducting one-dimensional edge contacts to the hBN-encapsulated graphene channel. Finally, air bridge contacts to inner SC electrodes were defined using the same process outlined above.

\subsubsection*{Measurement}
Measurements were primarily performed in a \textsuperscript{3}He cryostat (Oxford Instruments HelioxVL) with a base temperature of $250~\text{mK}$, equipped with extensive low-pass filtering using cryogenic $\pi$ and RC filters. We employed standard AC lock-in techniques with an excitation frequency of $13.333\text{--}43.33~\text{Hz}$ and an AC current of $1\text{--}10~\text{nA}$ (Stanford Research Systems SR830 and SR860). To generate the bias current, output voltages from DC sources (Yokogawa GS200) and the AC lock-in oscillator were converted into currents using independent series resistors (typically $1~\text{M}\Omega$ and $100~\text{M}\Omega$, respectively) and combined at a resistive summing node at room temperature. Quasi-four-probe voltage measurements were performed using a low-noise differential voltage preamplifier (Stanford Research Systems SR560), while the current was drained through a low-noise current preamplifier (DL Instruments Model 1211). DC voltage measurements were performed using a digital multimeter (Keysight 34401A) to probe zero-field superconducting characteristics. For the graphene control device, a global bottom gate voltage was applied to the heavily doped Si substrate using a source measure unit (Keithley 2400).

Crucially, field-dependent measurements in the CJJ devices were performed by thermally cycling the samples above the transition temperature of the SC electrodes, followed by field-cooling at each magnetic field point. The magnetic field was generated by a superconducting magnet driven by a DC current source (Yokogawa GS200). This procedure allows magnetic flux to penetrate the junction through the outer SC contact. For linear JJ devices, such field-cooling is unnecessary, as magnetic flux can enter continuously through the open boundaries. To evaluate the JDE, bias current sweeps were consistently performed outward from zero for both current directions. This ensures that the diode effect is determined by the switching current rather than the retrapping current.

\subsection*{Data availability}
The data that support the findings of this study are presented in the Article, Extended Data, and Supplementary Information. Further data are available from the corresponding authors upon reasonable request.


\end{document}


\title{Supplementary Information: Vortex-parity-controlled diode effect in Corbino topological Josephson junctions}
\maketitle

\subsection*{I. Surface protection of Sn-doped $\rm{Bi_{1.1}Sb_{0.9}Te_{2}S}$ crystal}

$\rm{Sn_{0.02}Bi_{1.08}Sb_{0.9}Te_{2}S}$ (Sn-BSTS) is a promising three-dimensional topological insulator (3DTI) known for its highly bulk-insulating character and surface-dominated transport~\cite{kushwaha2016a}. However, surface degradation remains a critical challenge. While air sensitivity is a known issue for the tetradymite 3DTI family such as $\rm{Bi_{2}Se_{3}}$~\cite{kong2011,salehi2015}, we observe that Sn-BSTS is particularly sensitive to air exposure. We attribute this extreme instability to its complex non-integer stoichiometry involving five elements, which likely creates a thermodynamic drive for rapid surface reaction with oxygen or moisture. Figure S1a displays a typical atomic force microscopy (AFM) topography image of a Sn-BSTS surface cleaved and exposed to ambient conditions. The surface is heavily textured with nanobubbles whose heights significantly exceed the single-quintuple-layer step height. We attribute the challenge in realizing high-quality Josephson junctions using Sn-BSTS to this surface instability, which compromises its exceptional properties upon minimal atmospheric exposure.

To address this, we adopted the surface protection strategy widely utilized in the field of molecular beam epitaxy (MBE) for growing high-quality chalcogenide 3DTI thin films (e.g., $\rm{(Bi_{\it{x}}Sb_{1-\it{x}})_{2}Te_{3}}$), where in situ Te capping is a well-established practice. By integrating an ultrahigh-purity Ar glovebox directly with an ultrahigh vacuum (UHV) MBE chamber, we extend this effective technique to mechanically exfoliated flakes, bridging the gap between micomechanical processing of van der Waals (vdW) layers and UHV MBE-grade surface preservation.

This approach is particularly well-suited for Sn-BSTS because the crystal cleaves at the vdW gap between adjacent Te atomic planes, providing a chemically compatible template~\cite{kushwaha2016a}. Furthermore, we leverage the advantageous growth properties of Te established in the MBE community: trigonal Te consists of helical 1D vdW chains, which enables strain-free vdW epitaxy on tetradymite surfaces~\cite{park2015,okuyama2017,richardson2017}. Finally, consistent with previous studies, the Te capping layer remains insulating at low temperatures, ensuring it protects the surface without shunting the transport~\cite{hoefer2015,liang2021}.

Figure S1b validates this approach. The AFM image shows a Sn-BSTS flake capped with $4~\text{nm}$ of Te immediately after exfoliation in the inert environment. In contrast to the air-exposed sample, the capped surface reveals clear atomic terraces of the underlying Sn-BSTS alongside the grain structure of the Te film, showing negligible signs of degradation. This confirms that applying the in vacuo Te capping technique to crystallite flakes exfoliated from a bulk single crystal is an effective pathway to preserve the intrinsic surface quality of Sn-BSTS for device fabrication.

\subsection*{II. Analogous multiple-slit interference model for Corbino-geometry Josephson junctions}

While the main text presents the formalism for the critical current modulation in Corbino-geometry Josephson junctions (CJJs), the mechanism behind the selection rules can be intuitively understood by mapping the polygonal geometry to an optical multiple-slit interference model. As schematically illustrated in Fig. 1b and Fig. S2, a regular polygon with $n_{\rm c}$ corners can be effectively modelled as a symmetric direct current (DC) superconducting quantum interference device (SQUID) comprising $n_{\rm c}$ junctions (sides) and $n_{\rm c}$ loops (corners). The corners introduce discrete phase shifts between sides due to their finite area, breaking the continuous rotational symmetry of a circular junction.

In this model, the magnetic flux $\Phi_{\rm JJ}$ threading the weak link induces a total winding of the gauge-invariant superconducting (SC) phase difference along the closed azimuthal perimeter of the junction. Unlike linear junctions, fluxoid quantization around this closed loop restricts the physically accessible states to those where the total phase winding is exactly $2\pi n_{\rm v}$ for an integer $n_{\rm v}$. This total phase is distributed evenly among the $n_{\rm c}$ segments (each segment comprising a side and a corner). Here, the vortex number corresponds to the normalized flux $n_{\rm v} = \Phi_{\rm JJ}/\Phi_0$, where $\Phi_0 = h/2e$ is the SC magnetic flux quantum ($h$ is the Planck constant and $e$ is the elementary charge). Constructive interference---analogous to principal maxima in optical diffraction---occurs only when the phase difference across each segment, equivalent to a ``slit", is a multiple of $2\pi$. This leads to the fundamental condition for a non-zero critical current: the vortex number $n_{\rm v}$ must satisfy $n_{\rm v}/n_{\rm c} = m$ (where $m$ is an integer).

Figure S2 illustrates this concept by comparing the continuous interference pattern (analogous to optical intensity) with the discrete sampling imposed by the Corbino geometry.
\begin{itemize}
    \item \textbf{Circular limit ($n_{\rm c} \to \infty$):} For a perfectly circular junction, the interference pattern approaches the Fraunhofer diffraction pattern of a single slit (Fig. S2a, green line). However, as defined by the fluxoid quantization, the accessible states are restricted to integer vortex numbers $n_{\rm v}$. In a homogeneous junction, these integer points align exactly with the nodes (zeros) of the Fraunhofer pattern for all $n_{\rm v} \neq 0$. Consequently, despite the theoretical presence of a diffraction envelope, the junction exhibits zero critical current at any finite field. Note that this complete destructive interference of Josephson currents persists for all harmonic numbers ($k \geq 1$) of the current-phase relation (CPR), as the node alignment condition is independent of $k$. Thus, the circular geometry inherently lacks any observable critical current at finite flux.
    \item \textbf{Square geometry ($n_{\rm c} = 4$):} The interference pattern corresponds to that of a 4-slit diffraction pattern (Fig. S2b). For a sinusoidal CPR dominated by the first harmonic ($k=1$), the principal maxima occur at $n_{\rm v} = 4m$, which are the only points where the discrete sampling (circles) coincides with a non-zero amplitude (blue curve).
    \item \textbf{Higher harmonics ($k=2$):} If the CPR contains a second harmonic component ($k=2$, representing Cooper pair cotunnelling), the phase periodicity of this specific term is effectively halved. In the diffraction picture, this is analogous to doubling the spatial frequency of the interference fringes. Consequently, constructive interference peaks for the $k=2$ component appear at all even integers ($n_{\rm v} = 2m$). Crucially, at vortex numbers satisfying $n_{\rm v} = 4m + 2$ (e.g., $n_{\rm v} = \pm 2$), the dominant first-harmonic contribution vanishes due to destructive interference. Therefore, the observation of finite supercurrents at these specific points serves as a direct signature of the higher-harmonic content. By selectively filtering harmonic components based on the vortex number, the square geometry---and its generalization to arbitrary polygons---effectively performs a Fourier decomposition of the CPR through simple DC transport measurements in a single JJ device.
\end{itemize}

\subsection*{III. Superconducting properties and interface transparency of 3DTI CJJs}

From the dependence of the junction differential resistance on temperature $T$ (Extended Data Fig. 2a), we extract the junction critical temperatures $T_{\rm c,JJ} = 3.0~\text{K}$ and $3.3~\text{K}$ for the circular and square 3DTI CJJs, respectively. Two additional resistive transitions are observed at higher temperatures, which we attribute to the superconducting transitions of both the proximitized Sn-BSTS underneath the Nb electrodes and the Nb electrodes themselves. A detailed analysis of these energy scales involves extracting the induced gap, as discussed below.

In differential conductance measurements, we observe finite-voltage features consistent with multiple Andreev reflections (MAR; Fig. S3a,b). To extract the effective gap, we adopt the analysis described by Schmitt et al.~\cite{schmitt2022}. The nonlinear conductance features occur at voltages corresponding to subharmonics of the induced gap, following the relationship $eV = 2\Delta^*(T)/n$ for integer $n$, where $V$ is the voltage across the junction. We fit the temperature dependence of the induced gap $\Delta^*(T)$ using the following model:
\begin{equation}
    \Delta^*(T)=\frac{\Delta_{\rm BCS}(T)}{1+\gamma_{\rm B}\sqrt{\Delta_{\rm BCS}(T)^2-\Delta^*(T)^2}/\pi k_{\rm B}T_{\rm c}},
\end{equation}
where $\Delta_{\rm BCS}(T)=1.764k_{\rm B}T_{\rm c}\tanh(1.74\sqrt{T_{\rm c}/T-1})$ represents the Bardeen--Cooper--Schrieffer (BCS) parent gap of the superconductor, $k_{\rm B}$ is the Boltzmann constant, and $\gamma_{\rm B}$ parameterizes the interface transparency. The best-fit parameters are $T_{\rm c} = 7.1~\text{K}$, $\gamma_{\rm B}=1$, and $\Delta^*(0)=0.78~\text{meV}$ for the circular junction, and $T_{\rm c} = 7.3~\text{K}$, $\gamma_{\rm B}=0.9$, and $\Delta^*(0)=0.83~\text{meV}$ for the square junction. Notably, these extracted values for the parent and induced gaps are consistent with the resistive transitions observed in the resistance $R$ versus $T$ curves of these devices (Extended Data Fig. 2a).

We characterize the junction parameters—critical current $I_{\rm c}$, normal-state resistance $R_{\rm N}$, and excess current $I_{\rm e}$—from the DC voltage $V_{\rm{DC}}$ versus DC bias current $I_{\rm{DC}}$ characteristics measured at the base temperature of $T = 0.25~\text{K}$ (Fig. S3c,d; see also Extended Data Fig. 2). We determine $R_{\rm N}$ from the slope of the linear (ohmic) $I_{\rm{DC}}$--$V_{\rm{DC}}$ regime at high bias ($I_{\rm{DC}} \gg I_{\rm c}$), yielding $179~\Omega$ and $173~\Omega$ for the circular and square junctions, respectively. These values are consistent with the differential resistance saturation values observed at high temperatures ($181~\Omega$ and $174~\Omega$), where $T=10~\text{K} > T_{\rm{c,Nb}} \approx 7~\text{K}$ (see Extended Data Fig. 2a). Furthermore, we extract the excess current $I_{\rm e}$ by extrapolating the linear high-bias fit to zero voltage (Fig. S3c,d), obtaining $I_{\rm e} = 3.22~\mu\text{A}$ and $4.53~\mu\text{A}$ for the circular and square junctions, respectively. These values correspond to $eI_{\rm e}R_{\rm N}/\Delta^*$ ratios of $0.74$ and $0.94$. Using the Octavio--Tinkham--Blonder--Klapwijk (OTBK) theory~\cite{flensberg1988}, these ratios translate to transparencies of $\mathcal{T} = 0.68$ and $0.73$, respectively, demonstrating the high quality of the Josephson junctions fabricated on the surface-protected Sn-BSTS. Extracted parameters for both junction geometries are summarized in Extended Data Table 1.

\subsection*{IV. Flux periodicity and effective area of CJJs}

The discrete vortex entry events described in the main text are visualized as sharp discontinuities in the differential resistance waterfall plots shown in Fig. S4a--c. By tracking these resistance jumps as a function of the applied magnetic field $B$, we extract an average flux periodicity of $\Delta B = 1.28~\text{mT}$ and $1.33~\text{mT}$ for the circular and square 3DTI CJJs presented in the main text, respectively. These values correspond to effective flux-quantizing areas of $A_{\rm eff} \approx 1.62~\upmu\text{m}^2$ and $1.56~\upmu\text{m}^2$.

Defining the exact area responsible for fluxoid quantization in mesoscopic SC rings requires considering the magnetic field penetration into the superconductor. We define an effective area $A_{\rm eff}$ that encompasses the geometric area enclosed by the outer SC ring plus a correction due to an effective penetration depth $\lambda_{\rm eff}$. For the circular geometry, $A_{\text{circle,eff}}=\pi(r_{\text{out}}+\lambda_{\rm eff})^2$, and for the square geometry, $A_{\text{square,eff}}=(a_{\text{out}}+2\lambda_{\rm eff})^2$ (see schematics in Fig. S4d--f). By fitting the measured periodicities to $\Delta B = \Phi_0 / A_{\rm eff}$, we extract a consistent effective penetration depth of $\lambda_{\rm eff} \approx 181~\text{nm}$ for both the circular and square junctions. This value is comparable to the magnetic penetration depth reported for Nb films of similar thickness~\cite{gubin2005}.

To further validate this geometric model and the universality of the extracted penetration depth, we fabricated and measured a larger square junction as a control device, with a designed outer length roughly double that of the device in the main text, while maintaining the same junction length (Fig. S4c,f). The resistance jumps for this control device exhibit an average periodicity of $\Delta B = 0.629~\text{mT}$, corresponding to an effective area of $A_{\rm eff} \approx 3.29~\upmu\text{m}^2$. Applying the same area model ($A_{\text{square,eff}}$), we obtain $\lambda_{\rm eff} = 176~\text{nm}$. The remarkable consistency of $\lambda_{\rm eff}$ across three distinct devices with different geometries and sizes confirms that the flux periodicity in our Corbino CJJs is robustly determined by the area enclosed by the outer superconductor, extended by a characteristic field penetration length.

\subsection*{V. Robustness of even-odd diode effect in tight-binding model}

Here, we discuss the robustness of the calculated diode effect to changes of parameters in our tight-binding simulations. The theoretical model comprises four Majorana rings, two describing the inner edge and two describing the outer edge  \cite{lesser2025}. The model contains three hopping amplitudes: $t_0$ (intra-ring hopping), $t_1$ (direct inter-ring hopping within the outer / inner pair), and $t_2$ (``diagonal" inter-ring hopping within the outer / inner pair). We set $t_1=2t_2$ to get gapless chiral edge modes, however the ratio $t_1/t_0$ is a free parameter. The model also contains a parameter $\Delta$, which is the coefficient of a two-Majorana term between the inner and outer edges; the ratio $\Delta/t_0$ is a free parameter.

Main text Fig. 3 shows results for the diode effect for $t_1/t_0=0.6$ and $\Delta/t_0=0.2$. We have confirmed that changing these ratios by a factor of 2 (in both directions) leaves the qualitative picture unchanged: the diode effect exhibits even-odd polarity switching, though the magnitude of the diode effect depends on details. Likewise, in our tight-binding model the perfect rotation symmetry is broken by a single impurity (local change in the value of $\Delta\to\Delta'$). We have verified that the precise value of $\Delta'/\Delta$ also does not make a qualitative difference, within at least a factor of 2, and so does the spread of the impurity (from 2 sites and up to at least 10 sites without any qualitative change). Directly matching the model and experiment is challenging, so we focus on placing the model within a regime of parameters that supports our understanding of the edge physics. For example, taking $\Delta\gg t_0$ would not make sense because then talking about an ``inner" and an ``outer" edge would not be sensible, as they would become completely mixed. Likewise, small deviations from $t_1=2t_2$ are reasonable (leading to small gaps), but significant deviations would mean the edges are fully gapped, and this picture would not truly represent a good topological insulator with a large bulk gap. Our analysis shows that if we keep the model within this reasonable regime of parameters, the even-odd diode polarity switching persists, supporting the notion of its origin being topological.

\bibliography{main2}

\newpage

\begin{figure}
    \centering
    \includegraphics[width = \textwidth]{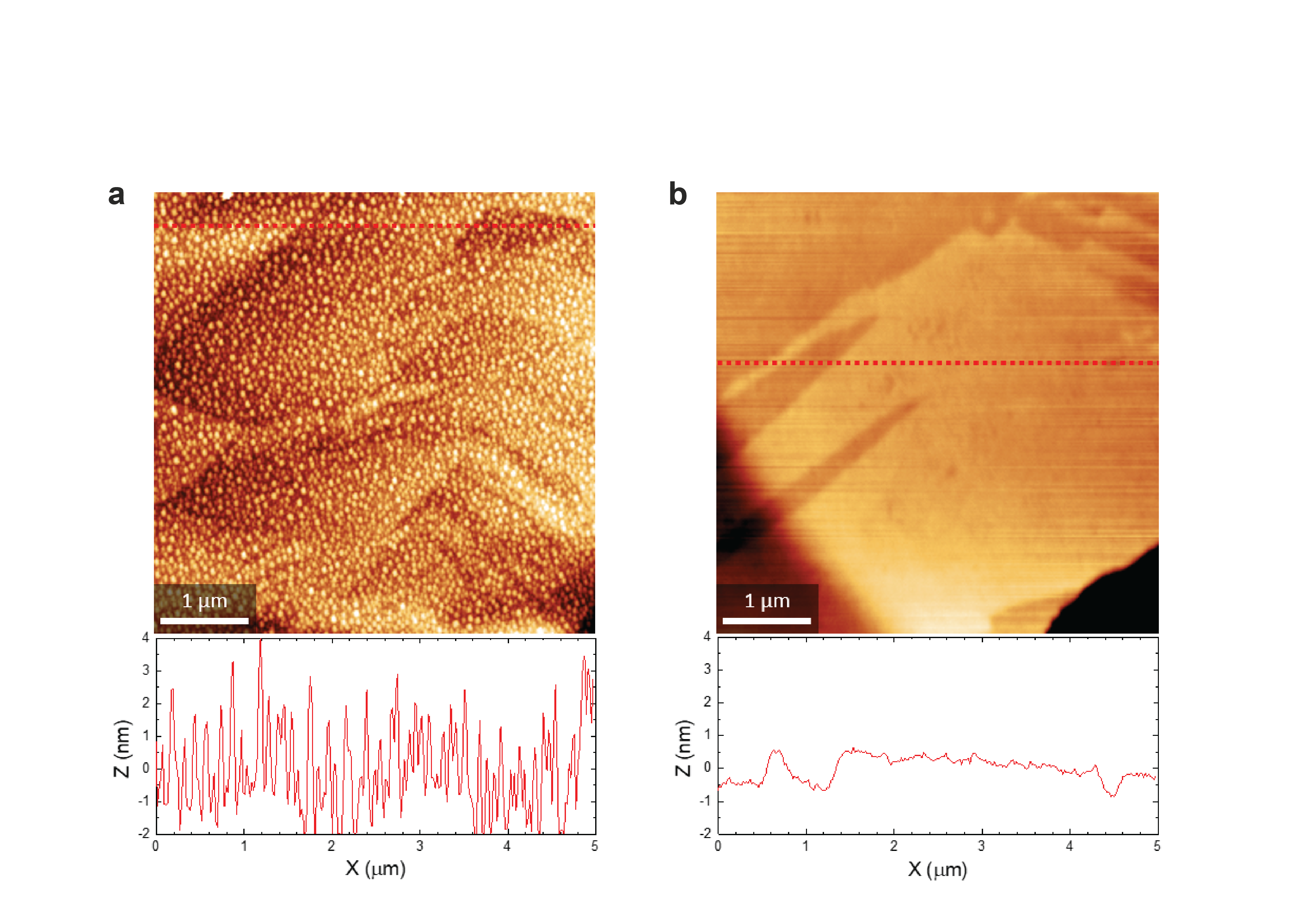}
    \caption{
    \textbf{Surface protection of Sn-BSTS single crystals.}
    \textbf{a}, AFM topography (top) and height profile along the red dotted line (bottom) of an uncapped Sn-BSTS surface after air exposure. The surface is heavily textured with nanobubbles, obscuring the underlying atomic steps.
    \textbf{b}, Corresponding data for a Sn-BSTS surface capped in vacuo with $4~\text{nm}$ of Te prior to air exposure. The atomic terraces of the underlying Sn-BSTS are clearly resolved along with the Te grain structure, demonstrating effective surface protection.
    }
\end{figure}

\begin{figure}
    \centering
    \includegraphics[width = 0.8 \textwidth]{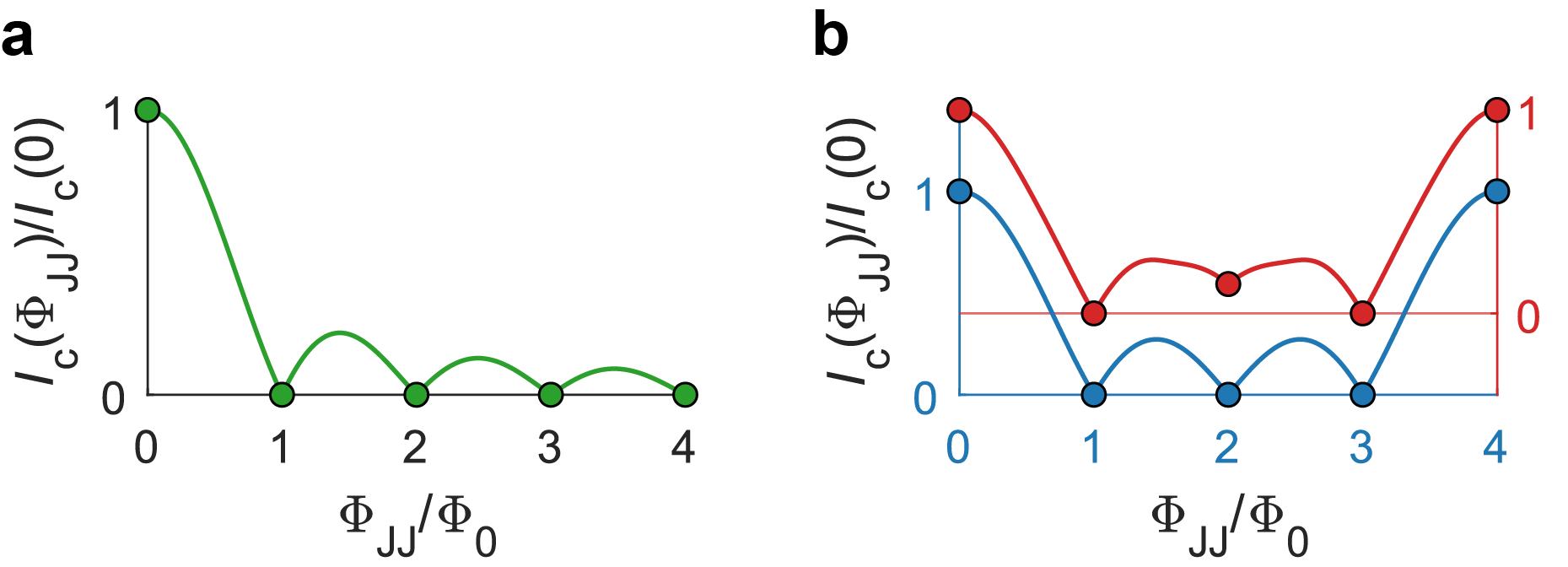}
    \caption{
    \textbf{Schematic of selection rules in Corbino--Josephson interferometry.}
    \textbf{a}, Interference model for a circular CJJ. The green line represents the Fraunhofer diffraction pattern (analogous to single-slit optics) calculated as a function of continuous normalized flux $\Phi_{\rm JJ}/\Phi_0$. The green circles indicate the physically accessible states, restricted by fluxoid quantization to integer vortex numbers $n_{\rm v}$ (where $\Phi_{\rm JJ}/\Phi_0 = n_{\rm v}$). Note that the allowed states align with the nodes of the diffraction pattern for all non-zero integers.
    \textbf{b}, Interference model for a square CJJ ($n_{\rm c}=4$), analogous to 4-slit optical interference. The blue curve shows the interference pattern for a purely sinusoidal ($k=1$) CPR plotted against $\Phi_{\rm JJ}/\Phi_0$, while the circles denote the allowed discrete states. Finite critical currents appear only when the vortex number $n_{\rm v}$ is a multiple of 4. The red curve and circles (vertically offset for clarity) illustrate the effect of a second-harmonic ($k=2$) component present in the CPR. While the dominant first harmonic vanishes at non-multiple-of-4 even integers ($n_{\rm v} = 4m + 2$), the $k=2$ component (having half the phase periodicity) contributes constructively, resulting in observable critical currents (node lifting) at these specific vortex numbers.
    }
\end{figure}

\begin{figure}
    \centering
    \includegraphics[width = \textwidth]{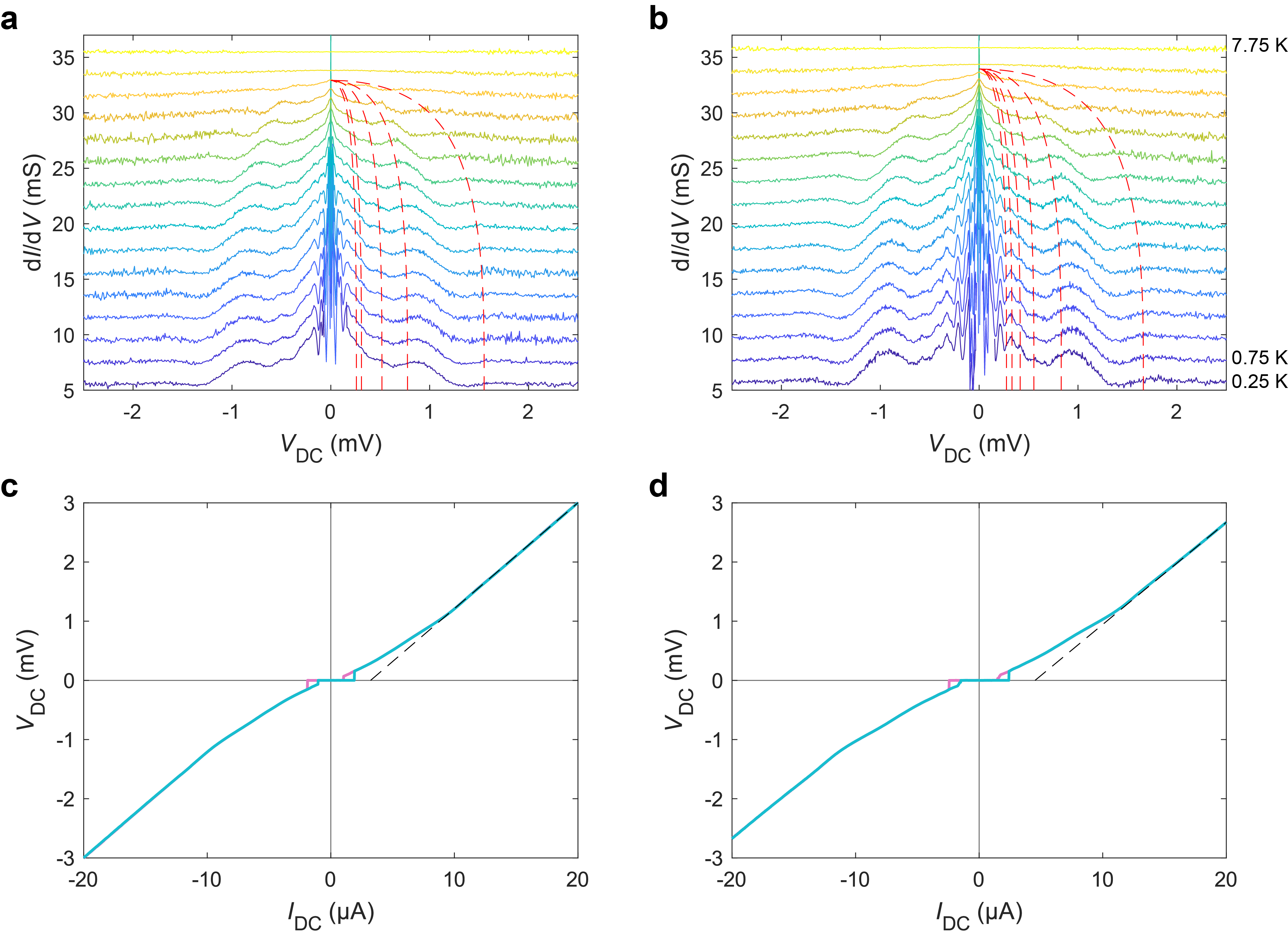}
    \caption{
    \textbf{Analysis of multiple Andreev reflections, normal-state resistance, and excess current.}
    \textbf{a}, Differential conductance $dI/dV$ versus DC voltage $V_{\rm DC}$ for the circular 3DTI CJJ at temperatures ranging from $0.25$ to $7.75~\text{K}$ (in $0.5~\text{K}$ steps). Curves are vertically offset by $2~\text{mS}$ for clarity. Broad conductance peaks corresponding to MAR are observed at voltages satisfying $eV_{\rm DC}=2\Delta^*(T)/n$. Dashed red lines represent the fit to Eq. (1) using the parameters described in the text for subharmonic orders $n=1,2,3,5,~\text{and}~6$.
    \textbf{b}, Corresponding data for the square 3DTI CJJ. Dashed red lines represent Eq. (1) for $n=1,2,3,4,5,~\text{and}~6$.
    \textbf{c}, DC voltage $V_{\rm DC}$ versus DC bias current $I_{\rm DC}$ for the circular junction at $T=0.25~\text{K}$. Forward and backward current sweeps are shown in blue and pink, respectively. The dashed line indicates a linear fit to the high-bias ohmic regime ($I_{\rm DC} \gg I_{\rm c}$), used to extract the excess current $I_{\rm e}$ (current-axis intercept at $V_{\rm DC}=0$) and the normal-state resistance $R_{\rm N}$ (slope).
    \textbf{d}, Corresponding $V_{\rm DC}$ versus $I_{\rm DC}$ data for the square junction.
    All data are measured after careful zero-field cooling.
    }
\end{figure}

\begin{figure}
    \centering
    \includegraphics[width = \textwidth]{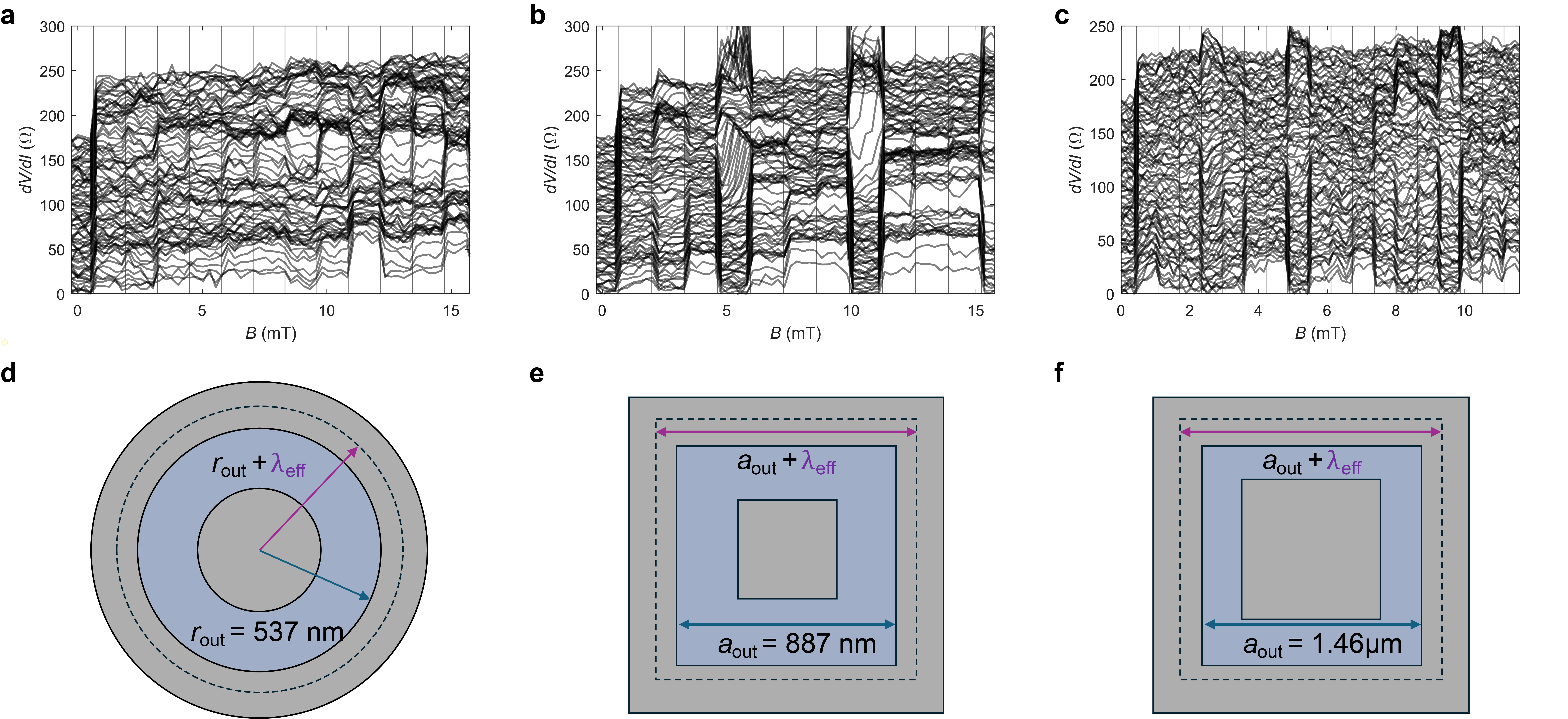}
    \caption{
    \textbf{Flux periodicity and extraction of effective penetration depth.}
    \textbf{a--c}, Waterfall plots of differential resistance versus applied magnetic field $B$ for the circular junction (\textbf{a}) and the square junction (\textbf{b}) presented in the main text, and for a larger square junction measured as a control device (\textbf{c}), taken at $T=1.6~\text{K}$. Each curve corresponds to a single DC bias current, vertically offset for clarity. Vertical lines indicate the average periodicity $\Delta B$ of the resistance jumps: $1.28~\text{mT}$ (\textbf{a}), $1.33~\text{mT}$ (\textbf{b}), and $0.629~\text{mT}$ (\textbf{c}).
    \textbf{d--f}, Schematics illustrating the nominal design dimensions and effective flux-quantizing areas for the corresponding devices. The dashed lines indicate the effective boundary extended by the effective penetration depth $\lambda_{\rm eff}$.
    \textbf{d}, Circular junction with a nominal outer radius $r_{\text{out}}=537~\text{nm}$. Fitting to $\Delta B = \Phi_0 / \pi(r_{\text{out}}+\lambda_{\rm eff})^2$ yields $\lambda_{\rm eff} = 181~\text{nm}$.
    \textbf{e}, Square junction with a nominal outer side length $a_{\text{out}}=887~\text{nm}$. Fitting to $\Delta B = \Phi_0 / (a_{\text{out}}+2\lambda_{\rm eff})^2$ yields $\lambda_{\rm eff} = 181~\text{nm}$.
    \textbf{f}, Larger square control device with $a_{\text{out}}=1.46~\upmu\text{m}$. The extracted $\lambda_{\rm eff} = 176~\text{nm}$ is consistent with the smaller devices, confirming the validity of the effective area model.
    }
\end{figure}


\title{Extended Data: Vortex-parity-controlled diode effect in Corbino topological Josephson junctions}
\maketitle

\begin{figure*}
    \centering
    \includegraphics[width = \textwidth]{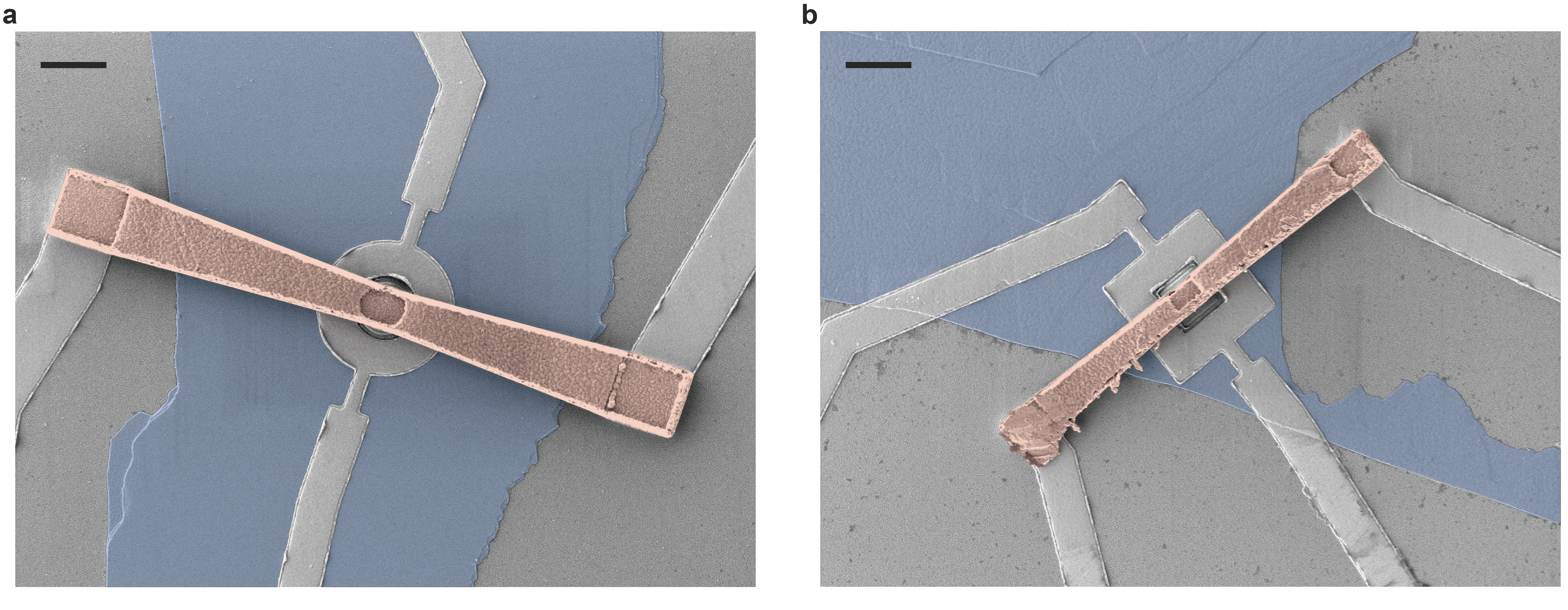}
    \newline
    \newline
    \justifying
    \textbf{Extended Data Fig. 1. Plan-view SEM images of the 3DTI CJJ devices.}
    False-colour SEM images of the circular (\textbf{a}) and square (\textbf{b}) devices. Scale bar: $1~\upmu \text{m}$. Au-capped Nb Corbino contacts and leads are coloured grey, the Au air bridges light red, and the Te-capped 3DTI flakes light blue. Note that the Te capping layer covers the entire substrate, including the $\text{SiO}_2/\text{Si}$ surface (uncoloured, appearing dark grey).
\end{figure*}

\begin{figure*}
    \centering
    \includegraphics[width = \textwidth]{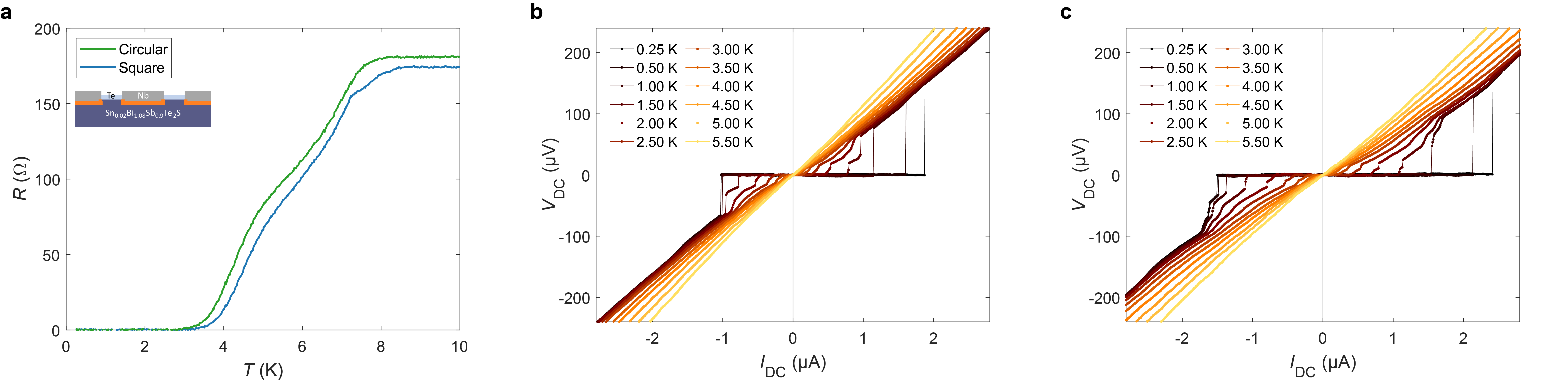}
    \newline
    \newline
    \justifying
    \textbf{Extended Data Fig. 2. Zero-field junction characteristics.} \textbf{a}, Temperature dependence of resistance for the two 3DTI CJJ devices discussed in the main text. Three distinct superconducting transitions are observed; from lowest to highest temperature, these are attributed to the junction, the proximitized 3DTI region underneath the Nb contacts, and the Nb contacts themselves. Inset: Schematic cross-section of the device structure (dark blue: 3DTI flake; light blue: Te capping layer; gray: Nb contacts; orange: proximitized 3DTI).
    \textbf{b},\textbf{c}, $I_{\rm{DC}}$--$V_{\rm{DC}}$ curves for the circular (\textbf{b}) and the square (\textbf{c}) 3DTI CJJ devices at various temperatures. Data were obtained by sweeping the DC bias current from negative to positive. Consequently, the transition to the zero-voltage state on the negative bias side corresponds to the retrapping current, while the transition to the resistive state on the positive bias side corresponds to the switching current.
\end{figure*}

{\renewcommand{\arraystretch}{1.8}
\begin{table}[h]
    \centering
    \begin{tabular}{|c|c|c|c|c|c|c|c|c|c|c|c|}
        \hline
        Device & $T_{\rm{c,JJ}}$ (K) & $R_{\rm{N}}$ ($\Omega$) & $I_{\rm{c}}$ ($\mu$A) & $I_{\rm{e}}$ ($\mu$A) & $eI_{\rm{c}}R_{\rm{N}}$ (meV) & $eI_{\rm{e}}R_{\rm{N}}$ (meV) & $\Delta_{\rm{Nb}}$ (meV) & $\Delta^*$ (meV) & $\frac{eI_{\rm{c}}R_{\rm{N}}}{\Delta^*}$ & $\frac{eI_{\rm{e}}R_{\rm{N}}}{\Delta^*}$ & $\mathcal{T}$ \\
        \hline
        Circluar & 3.0 & 179 & 1.91 & 3.22 & 0.342 & 0.576 & 1.1 (7.1 K) & 0.78 (5.1 K) & 0.44 & 0.74 & 0.68 \\
        \hline
        Square & 3.3 & 173 & 2.40 & 4.53 & 0.415 & 0.781 & 1.1 (7.3 K) & 0.83 (5.5 K) & 0.50 & 0.94 & 0.73 \\
        \hline
    \end{tabular}
    \newline
    \newline
    \textbf{Extended Data Table 1. Junction parameters}, extracted as described in SI Section III.
    \label{tab:junction_parameters}
\end{table}
}
\begin{figure*}
    \centering
    \includegraphics[width = \textwidth]{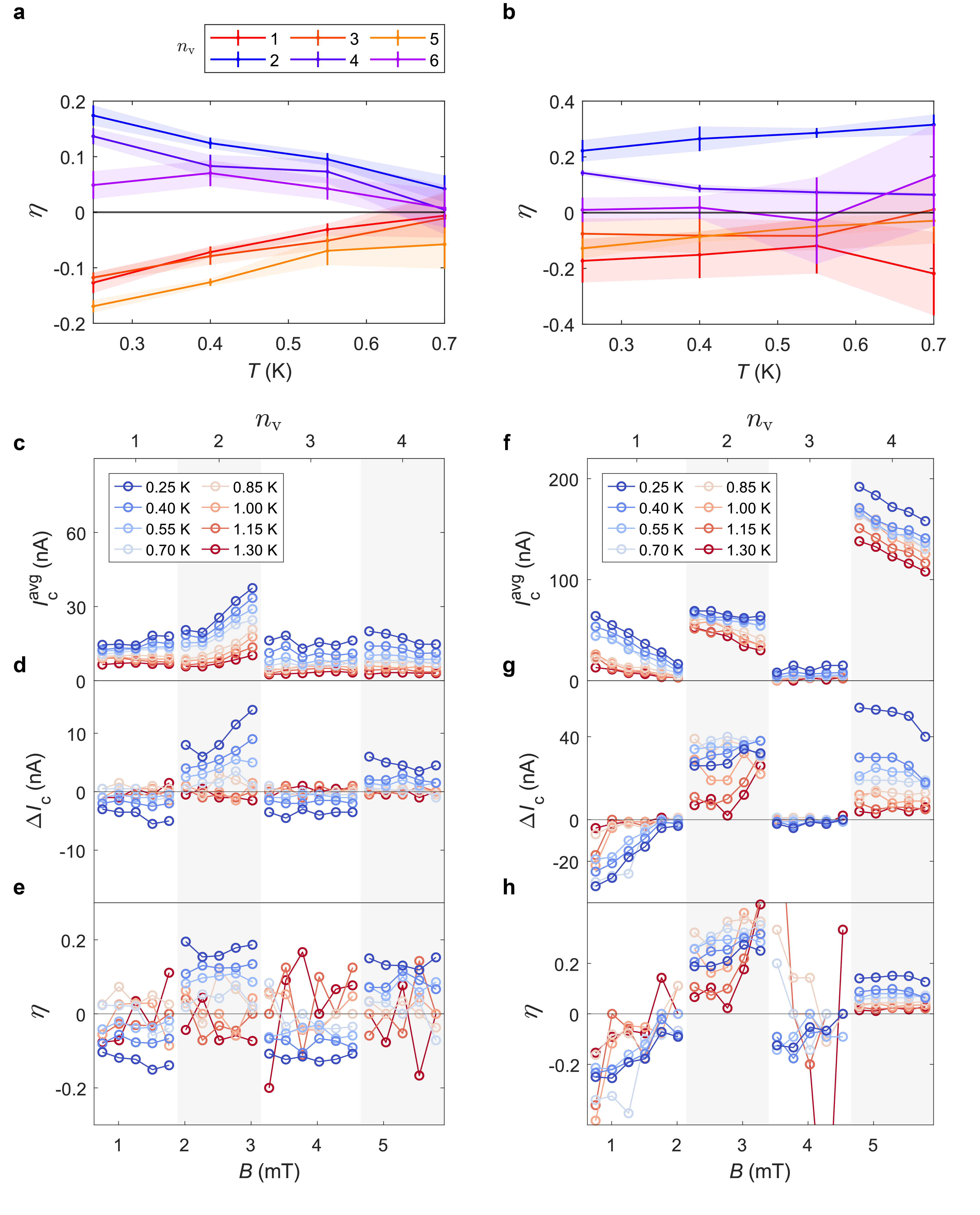}
    \newline
    \newline
    \justifying
    \textbf{Extended Data Fig. 3. Temperature dependence of the vortex-parity effect.}
    \textbf{a}, Diode efficiency of the circular 3DTI CJJ as a function of temperature for indicated vortex numbers $n_{\rm v}$. Error bars represent the standard deviation of data across different applied magnetic fields within each fixed $n_{\rm v}$ state; solid lines connect the mean values at each temperature.
    \textbf{b}, Same as \textbf{a} for the square junction.
    \textbf{c}--\textbf{e}, Circular junction parameters versus $B$ at various temperatures: 
    \textbf{c}, Average critical current $I_{\rm c}^{\rm avg}(B)$.
    \textbf{d}, Critical current difference $\Delta I_{\rm c}(B)$.
    \textbf{e}, Diode efficiency $\eta(B)$.
    \textbf{f}--\textbf{h}, Corresponding data for the square junction.
    In contrast to the main text, current values in \textbf{c}, \textbf{d}, \textbf{f}, and \textbf{g} are not normalized by $I_{\rm c}(0)$ to explicitly show the thermal decay of the critical current.
\end{figure*}

\begin{figure*}
    \centering
    \includegraphics[width = \textwidth]{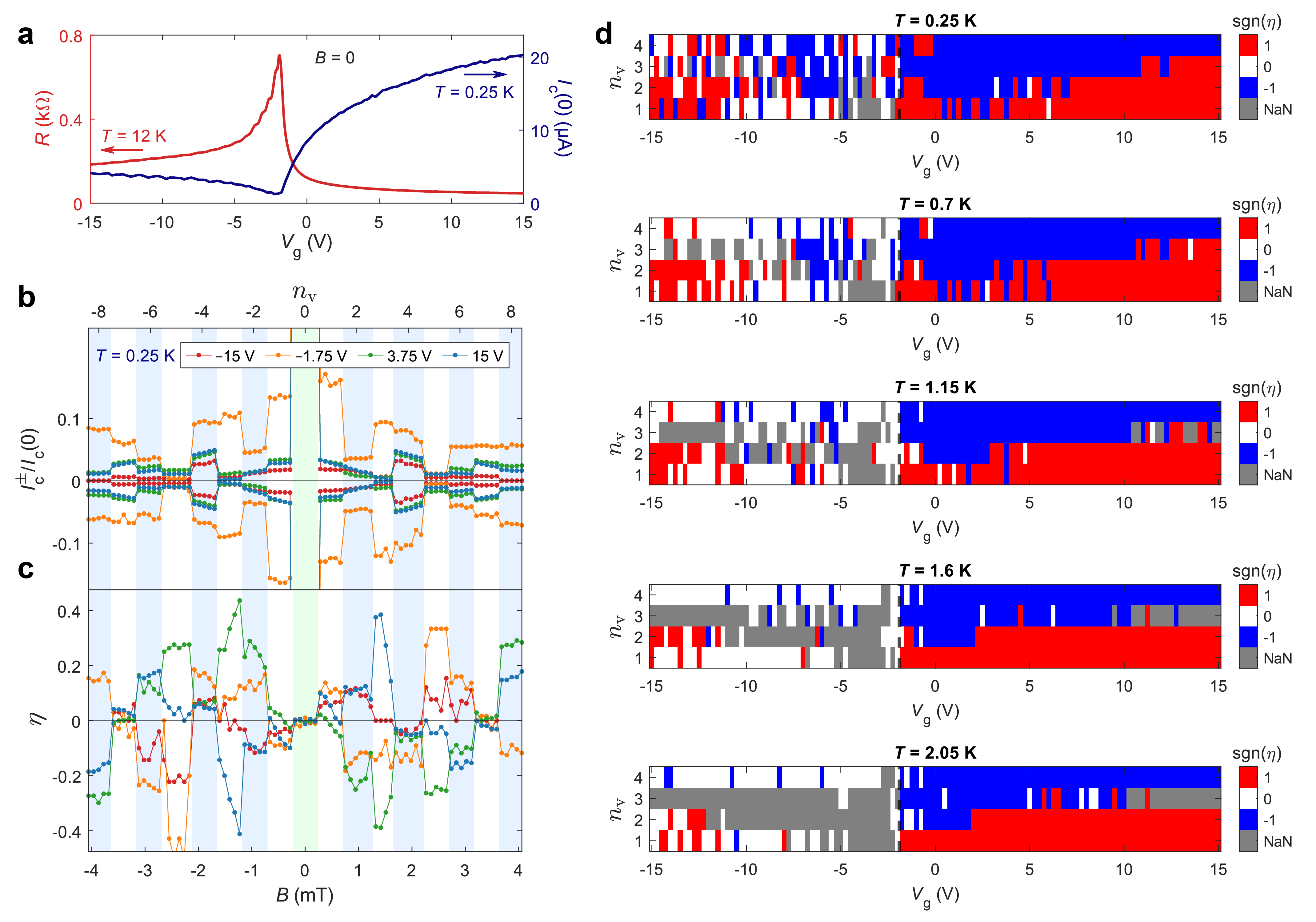}
    \newline
    \newline
    \justifying
    \textbf{Extended Data Fig. 4. Absence of vortex-parity-controlled JDE in graphene CJJ across an extensive parameter space.}
    \textbf{a}, Normal state resistance (measured at $T > T_{\rm{c,MoRe}}$) and base-temperature critical current of the graphene control device as a function of gate voltage $V_{\rm{g}}$, taken at $B=0$. The Dirac point is characterized by the resistance peak and the critical current dip at $V_{\rm{g}} \approx -1.9~\text{V}$.
    \textbf{b}, Normalized forward and backward critical currents versus magnetic flux for several gate voltages at $T=0.25~\text{K}$.
    \textbf{c}, Josephson diode efficiency $\eta$ derived from \textbf{b}. No systematic alternation with vortex number is observed, and $\eta$ varies irregularly with gate voltage for a fixed flux.
    \textbf{d}, Josephson diode polarity as a function of gate voltage and vortex number at various temperatures. Red (blue) indicates positive (negative) $\eta$. White represents data points with finite critical current but vanishing diode effect ($\Delta I_{\rm{c}} = 0$), while grey represents the resistive state ($I_{\rm{c}}^{\rm{avg}} = 0$). Throughout the explored parameter space, no vortex-parity-dependent sign alternation is observed. Data shown in the main text corresponds to $V_{\rm{g}} = 15~\text{V}$.
\end{figure*}